# Convection in Binary Fluid Mixtures.
# II. Localized Traveling Waves.

W. Barten[*] and M. Lücke

*Institut für Theoretische Physik, Universität des Saarlandes, D-66041 Saarbrücken, Germany*

M. Kamps

*HLRZ, Forschungszentrum, D-52425 Jülich, Germany*

R. Schmitz

*Institut für Festkörperforschung, Forschungszentrum, D-52425 Jülich, Germany*

Nonlinear, spatially localized structures of traveling convection rolls that are surrounded by quiescent fluid in horizontal layers of binary fluids heated from below are investigated in quantitative detail as a function of Rayleigh number for two different Soret coupling strengths (separation ratios) with Lewis and Prandtl numbers characterizing ethanol-water mixtures. A finite-difference method was used to solve the full hydrodynamic field equations numerically in a vertical cross-section perpendicular to the roll axes subject to realistic horizontal and laterally periodic boundary conditions with different periodicity lengths. Structure and dynamics of these localized traveling waves (LTW) are dominated by the concentration field. Like in the spatially extended convective states that are investigated in an accompanying paper, the Soret-induced concentration variations strongly influence, via density changes, the buoyancy forces that drive convection. The spatio-temporal properties of this feed-back mechanism, involving boundary layers and concentration plumes, show that LTW's are strongly nonlinear states. Light intensity distributions are determined that can be observed in side-view shadowgraphs done with horizontal light along the roll axes. Detailed analyses of all fields are made using colour-coded isoplots, among others. In the frame comoving with their drift velocity, LTW's display a nontrivial spatio-temporal symmetry consisting of time-translation by half an oscillation period combined with vertical reflection through the horizontal midplane of the layer. A time-averaged concentration current is driven by a phase difference between the waves of concentration and vertical velocity in the bulk of the LTW state. The associated large-scale concentration redistribution stabilizes the LTW and controls its drift velocity into the quiescent fluid by generating a buoyancy-reducing concentration "barrier" ahead of the leading LTW front. All considered LTW's drift very slowly into the direction of the phase velocity of the pattern. For weak Soret coupling, $\psi = -0.08$, LTW's have a small selected width and exist in a narrow band of Rayleigh numbers above the stability threshold for growth of TW's. For stronger coupling, $\psi = -0.25$, LTW's exist below the bifurcation threshold for extended TW's in a narrow band of Rayleigh numbers. In its lower part, LTW's have a small selected width. For somewhat higher Rayleigh numbers, there exist two LTW attractors with two different widths. For yet higher Rayleigh numbers, there is again only one LTW attractor, however, with a broader width. Dynamical properties and the dependence on the system length are analysed. Comparisons with experiments are presented.

47.20.-k,47.54.+r,47.15.-x,47.10.+g

[*]Present address: Paul Scherrer Institut, CH-5232 Villigen PSI, Switzerland

# I. INTRODUCTION

In this paper we investigate, by numerical integration of the hydrodynamic field equations, localized traveling wave (LTW) convection in binary fluid layers heated from below, i.e., convective structures that are spatially confined and surrounded by quiescent fluid. The convectively active region consists of a few straight, parallel roll vortices traveling under an intensity envelope which itself drifts, albeit very slowly compared to the phase velocity of the roll pattern. So a particular roll vortex grows under the trailing front of the LTW intensity envelope and travels towards the leading front, where its intensity smoothly decreases to zero.

Since a recent review [1] covers many of the experimental LTW observations [2–25] and many of the various theoretical approaches [26–45] capturing to a varying degree the experimental observations, we shall add only a few complementary comments – see also [26,27]. The fact that LTW structures of roll vortices do not exist in one-component fluids like water already suggests that the degrees of freedom of the concentration field play a decisive role: Firstly, they cause roll vortices to propagate when the separation ratio $\psi$ [1], i.e., the Soret coupling between temperature and concentration field, is negative and of sufficient size, whereas in pure fluids the primary convective structures are stationary. Secondly, traveling-roll vortices generate a large-scale concentration current. In LTW states, it causes a characteristic large-scale concentration redistribution that influences, via its contribution to the local density, the driving profile of the local buoyancy force. The balance of the latter turns out to be dominated by the current-induced concentration redistribution and to be of a form that (i) stabilizes the convectively active region against invasion of the quiescent conductive state and vice versa and (ii) controls in a delicate balance the drift of the LTW into the conductive state to be very slow by a build-up of a concentration "barrier" which impedes the LTW motion.

Our investigation of LTW's is based upon results of an analysis [46] of spatially-extended convection structures of stationary rolls, so-called SOC states, and of traveling rolls, so-called TW states, and, simultaneously, it is a continuation of this analysis. We have calculated and analysed several LTW states for two negative Soret coupling parameters, $\psi = -0.08$ and $\psi = -0.25$, that have also been investigated in experiments [2–15]. Here, the stronger Soret coupling $\psi = -0.25$ causes richer and more complex LTW bifurcation behaviour than the smaller one, $\psi = -0.08$. We should like to stress that all these LTW's are strongly nonlinear. Their spatiotemporal behaviour can neither be described quantitatively by a weakly nonlinear expansion around the oscillatory bifurcation threshold $r_{osc}$ for onset of TW convection, nor by simple heuristic amplitude-equation approaches. We therefore solved the full hydrodynamic field equations numerically with realistic boundary conditions, using an explicit finite-differences method. We restricted ourselves to 2d convection in the form of straight rolls by suppressing spatial variations along the roll axes. We imposed laterally-periodic boundary conditions with a periodicity length of twenty or forty times the height $d$ of the fluid layer. In particular, we wanted to elucidate the role of the concentration field. Without Soret coupling, $\psi = 0$, concentration deviations from the mean eventually diffuse away. It is the combination (i) of the linear coupling of its degrees of freedom to the temperature field via the Soret effect and (ii) of its contribution to the buoyancy force — the concentration field directly influences the driving mechanism for convection — that causes the additional richness of convective dynamics and structures in mixtures in comparison to pure one-component fluids. On the one hand, concentration is advected, and since the concentration diffusion constant is small, it is transported almost passively by the flow, except for boundary-layer effects. But, on the other hand, the structural dynamics of the concentration field actively feeds back via the associated density variations into the buoyancy force that drives convection.

LTW's have been investigated so far only in ethanol-water mixtures. They were found close to the oscillatory threshold $r_{osc}$ for sufficiently negative $\psi$. After the first observations of LTW's in short rectangular convection cells by Moses et al. [2] and Heinrichs et al. [3] for $\psi = -0.08$, Kolodner et al. [4,5] also saw, for $\psi = -0.25$, in an annular channel, i.e., a system uninfluenced by lateral endwalls, spatially-confined regions of TW activity surrounded by the quiescent conductive state. The convectively active region was stable, did not drift, and could have practically any width in different experimental runs for the same parameters. Then, in an ad hoc fifth-order amplitude equation, Thual and Fauve [28] found pulse solutions with selected width that drifted with the critical group velocity $v_g^{lin}$ of a linear wavepacket of TW perturbations at the oscillatory threshold. Furthermore, they showed that stability of these pulse solutions required complex coefficients. Also, the connection to soliton solutions of the nonlinear Schrödinger equation with imaginary coefficients was discussed [29–32]. Bestehorn et al. [33–36,47] investigated localized structures using their order-parameter equation. Van Saarloos and Hohenberg [37,38] found, for their amplitude equation, an analytical unstable pulse solution of selected width with a shape similar to a stable one. While all these pulses drift with $v_g^{lin}$, it was shown [39,40] that nonlinear gradient terms change drift velocity and pulse shape. With additional nonlinear terms in their amplitude equation, Levine and Rappel [41] reported solutions of arbitrary width and argued that the appearence of selected or arbitrary-width pulses sensitively depends on the coefficients in their equation. These simple amplitude-equation models have been very useful in providing a language to express certain aspects of the real LTW states. But they do not seem to be the appropriate theoretical framework for an explanation and a quantitative description of the physical phenomena appearing in the spatio-temporal behaviour of LTW's.

Niemela at al. [6] observed nominally the same LTW states with selected width for $\psi = -0.08$ in annular channels as well as in straight channels with endwalls. Their top-view shadowgraph intensity profiles seen with vertical illumination were similar to analytical pulse profiles of [37,38]. The LTW's in the annular channel came to rest after a transient drift with a positive or negative drift velocity that was small compared to the phase velocity $v_p$ of the roll structure. Furthermore, LTW's were generated [16,17] in finite, straight channels that rested in the middle and not near an endwall. Note that LTW's touching a lateral endwall change their structure slightly but their frequency strongly. For more details, see [17]. Anderson and Behringer [7,8] found LTW's of selected width above $r_{osc}$ in an annular cell. They also observed complex, transient, long-time behavior in their convection cells. With a photochromic technique, similar to the method of Moses and Steinberg [48,49], Katoh and Sawada [18] investigated the Lagrangian motion of a contaminant in LTW's of selected width in an annular channel.

Reference [42] then determined LTW states of the full 2d hydrodynamic field equations for $\psi = -0.25$ and $\psi = -0.08$ and ethanol-water parameters for one value $r$ of the reduced Rayleigh number, each in good agreement with the experiments [4–6,9]. The LTW's exhibited a strong similarity in their structure and their intrinsic mechanisms. Their group velocities were very small, $v_g \ll v_p$, but finite and positive. The structure of the concentration field showed that the LTW's were definitely not weakly-nonlinear structures – the concentration Péclet number was of the order of 1000. (Whether LTW's at smaller Soret coupling strength observed [17,19,20] in small rectangular cells are weakly nonlinear depends again on the structure of the concentration field or the size of the Péclet number.) A time-averaged circulation and redistribution of concentration over the entire extension of the LTW states was found [42]. It was related to the lateral mean concentration current in extended TW's [50,51]. The influence of the concentration current and the time-averaged concentration field via the time-averaged local buoyancy force on the group velocity $v_g$ of the LTW states was discussed [42,26]. The connection of the stability of the LTW states for $\psi = -0.08$ above $r_{osc}$ with the finite periodicity length of the system and the *convective* nature of the instability of the conductive state was mentioned; see also [21,22,52,53]. A comparison for $\psi = -0.25$ [9] and for $\psi = -0.08$ [10] showed quantitative agreement in the lateral structure of the numerically calculated [42] and the experimentally measured top-view shadowgraph intensity profiles of LTW's. Winkler and Kolodner [23] also discussed side-view shadowgraph intensities of a kind of transient LTW structure in a finite, rectangular cell. However, quantitative extraction of the LTW concentration field from experimental shadowgraphs, being more difficult than for extended TW's [23], has not yet been reported.

Kolodner [11] observed, in a thermally and geometrically more homogeneous [11] annular channel, that LTW's at $\psi = -0.08$ have the predicted [42] small but finite drift velocity. Furthermore, he found [10] that $v_g$ increases with $r$ and is mostly positive, i.e., in the same direction as $v_p$. Such an $r$-dependence was also seen by Ahlers [16] for slowly decelerating transients. At $\psi = -0.123$, LTW's at the lower end of the $r$-band of LTW stability had negative drift velocity [10]. Kolodner [10] used this fact to investigate the collision of LTW's. He found [12] different interaction behaviour than that reported by Brand and Deissler [40] for coupled fifth-order amplitude equations. Yahata [43] calculated and analysed, with a 2d MAC method [54], LTW's for the parameters of Katoh and Sawada [18], i.e., $\psi = -0.12$, in a finite cell with rigid lateral boundaries. His LTW's touch a lateral wall with their leading front. Various snapshots reveal a structure, also for the concentration field, that is similar to that of the drifting pulses of Ref. [42], despite of the different lateral boundary conditions.

Riecke [44,45] developed a kind of amplitude-equation system that describes a LTW not by just one single complex amplitude. He used an expansion for small Lewis numbers and found analytically, for idealized permeable boundary conditions, that he needed, in addition to the critical mode, a $x$ independent concentration mode that enters into the effective local Rayleigh number. The model of Cross [55–57] for counterpropagating TW's was thereby extended from four to five coupled partial differential equations. Compared to the pulse of Thual and Fauve [28,29], the critical mode is practically unaffected, but the drift velocity of the pulse is strongly reduced [44,45]. The behaviour of the concentration mode shows similarities [44] to the time-averaged concentration field obtained in the simulations of Ref. [42].

For a detailed discussion of the most recent developments concerning numerical solutions of the full field equations [27] and experiments [14,15], we refer the reader to the main part of this work; in particular secs. IV, V where our results are presented in quantitative comparison with recent experiments. Our work is organized as follows: In the accompanying paper [46], we investigate spatially-extended convection as the basis for understanding the more complicated LTW structures. Section II describes the system we study. In section III, the common structure of LTW states is discussed, especially focussing on the concentration field, its influence on the buoyancy balance, and the transport properties of LTW's. A comparison with extended TW convection is given. In section IV, we discuss bifurcation behaviour and the structural dynamics of narrow LTW's for $\psi = -0.08$. Section V investigates the dynamics and the more complicated bifurcation behavior of LTW's for $\psi = -0.25$, including finite-size effects. Special emphasis is given to the selection of width of these states. A detailed comparison with recent experiments is presented.

The conclusion in section VI lists our main results on strongly nonlinear LTW states.

## II. THE SYSTEM

We consider a horizontal layer of a binary fluid mixture like alcohol-water under a homogeneous gravitational field, $\mathbf{g} = -g\, \mathbf{e}_z$, that is directed downwards. A positive temperature difference $\Delta \underline{T}$ between the lower and upper confining boundaries is imposed externally, e.g., via high-thermal-conductivity plates in experiments. Here, we consider convection in the form of straight, parallel rolls, as seen in many experiments. Ignoring variations along the roll axes, we investigate 2d convection in an $x-z$ plane perpendicular to the axes described by the balance equations for mass, heat, concentration, and momentum in Oberbeck-Boussinesq approximation [58–61] as documented in [46, eq. (2.1-2.4)]. Lengths are scaled with the height $d$ of the layer, time with the vertical thermal diffusion time $d^2/\kappa$, and the velocity field $\mathbf{u} = (u, 0, w)$ by $\kappa/d$. Here, $\kappa$ is the thermal diffusivity of the mixture. $\delta T = (\underline{T} - \underline{T}_0)/\Delta \underline{T}$ denotes the deviation of the temperature from the mean temperature $\underline{T}_0$ in the fluid, scaled by the temperature difference between the plates. Some of the unscaled quantities are underlined for distinction from the scaled ones. The field $\delta C = (\underline{C} - \underline{C}_0)\, \beta / (\alpha\, \Delta\, \underline{T})$ is the scaled deviation of the concentration $\underline{C}$ from its mean $\underline{C}_0$. The thermal expansion coefficient $\alpha$ and the solutal expansion coefficient $\beta$ of the fluid come from a linear equation of state of the mass density $\underline{\rho} = \underline{\rho}_0 [1 - \alpha(\underline{T} - \underline{T}_0) - \beta(\underline{C} - \underline{C}_0)]$ for small deviations from the mean values $\underline{T}_0$ and $\underline{C}_0$.

Assuming the fluid to be incompressible the velocity field $\mathbf{u}$ is divergence-free. The reduced heat current $\mathbf{Q}$ consists of the convective part $\mathbf{u}\,\delta T$ and the diffusive part $-\boldsymbol{\nabla}\,\delta T$. In the reduced concentration current $\mathbf{J} = \mathbf{u}\,\delta C - L \boldsymbol{\nabla}\,(\delta C - \psi\,\delta T)$ the diffusive part is $-L \boldsymbol{\nabla}\,(\delta C - \psi\,\delta T)$. Here, we suppress the convective transport $\mathbf{u}\,F_0$ of the mean quantity $F_0$, since it drops out in the balance equations. The Lewis number $L = \frac{D}{\kappa}$ is the ratio of the concentration diffusivity $D$ to the thermal diffusivity $\kappa$, and the Prandtl number $\sigma = \frac{\nu}{\kappa}$ is the ratio of the momentum diffusivity $\nu$ and $\kappa$. For room temperatures ($10^\circ C - 40^\circ C$), the Prandtl number of ethanol-water mixtures lies between 5 and 20 [62]. The Lewis numbers are around 0.01. All our LTW calculations were done for $L = 0.01$ and $\sigma = 10$. The Rayleigh number $R = \frac{\alpha g d^3}{\kappa \nu}\Delta \underline{T}$ and the separation ratio $\psi = -\frac{\beta}{\alpha}\frac{k_T}{\underline{T}_0}$ are considered to be control parameters that can be varied independently. $R$ measures the externally imposed thermal stress. The Soret coupling $\psi$ between temperature and concentration, into which enters the thermo diffusion ratio $k_T$ of the mixture, reflects the influence of temperature gradients on the concentration field. For room temperature ethanol-water mixtures, $\psi$-values between about $-0.5$ and $+0.2$ can be easily realized experimentally [62].

The buoyancy force $(\rho - \rho_0)\mathbf{g}$ due to density deviations from the mean is the driving mechanism for convective motion. It enters into the momentum balance [46, eq. (2.4)] via the buoyancy term $\mathbf{B} = \sigma R\, (\delta T + \delta C)\mathbf{e}_z$. Taking the divergence of the momentum equation [46, eq. (2.4)], one gets, via the continuity equation, a Poisson equation [46, eq. (2.8)] for the pressure $p$ which builds together with [46, eq. (2.2-2.4)] a complete set of equations for the fields $\mathbf{u}$, $\delta T$, $\delta C$, and $p$.

The horizontal boundaries of the layer, that we shall call plates for shortness, are at $z = 0, 1$. The lateral boundaries are at $x = 0, \Gamma$. The plates are rigid for the fluid, and perfect heat conductors, so that the temperature of the fluid at $z = 0, 1$ is constant and laterally homogeneous. The plates are impermeable to the fluid, so there is no concentration current through the plates: $\mathbf{J} \cdot \mathbf{e}_z = 0$ or $\partial_z \delta C = \psi \partial_z \delta T$ at $z = 0, 1$. These NSI (no-slip impermeable) boundary conditions have to be contrasted with the idealized FSP (free-slip permeable) boundary conditions that are suited for expanding the fields in trigonometric functions and that are often used in theoretical approaches. However, since the latter allow vertical concentration transport through the plates, they probably misrepresent the delicate concentration balance of LTW states.

In lateral direction, all fields $F = u, w, T, C, p$ are periodic with a given lateral periodicity length $\Gamma$: $F(x, z; t) = F(x + \Gamma, z; t)$. Since the pressure $p$ is determined via the Poisson equation by $\mathbf{u}$, $\delta T$, $\delta C$, we do not need boundary conditions for it. Here, we report the results of calculations done in periodicity intervals of $\Gamma = 20$ and $40$. To integrate the partial differential equations we used a modification of the SOLA code that is based on the MAC method [63–66]. This is a finite-differences method of second order in space on staggered grids for the different fields, with an explicit first-order Euler step in time. The pressure field was iteratively calculated from the Poisson equation using the artificial viscosity method [66]. Calculations were done with uniform spatial resolution $\Delta x = \Delta z = 0.05$ on a CRAY. For more details see [27].

The conductive state $\mathbf{u}_{cond} = \mathbf{0}$; $\delta T_{cond} = \frac{1}{2} - z$; $\delta C_{cond} = \psi \left(\frac{1}{2} - z\right)$, is globally stable for small thermal stress. It describes vertical heat diffusion through the layer without convective motion. The heat current is $\mathbf{Q}_{cond} = \mathbf{e}_z$. The constant temperature gradient induces a vertical concentration gradient given by $-\psi$ which enforces the concentration current to vanish: $\mathbf{J}_{cond} = \mathbf{0}$. The concentration stratification causes a modification by the factor $(1 + \psi)$ in the buoyancy $\mathbf{B}_{cond} = \sigma R\, (1 + \psi) \left(\frac{1}{2} - z\right) \mathbf{e}_z$, relative to the thermal part. So, depending on the sign of the Soret coupling, the Soret effect enhances or depresses the buoyancy.

To describe convection, we use the velocity field, in particular the vertical component $w$ and the deviations from the conductive state: $\theta = \delta T - \delta T_{cond}$; $c = \delta C -$

$\delta C_{cond}$, $\mathbf{b} = \mathbf{B} - \mathbf{B}_{cond} = b\,\mathbf{e}_z$; $b = \theta + c$. A global measure of convection is $N-1$, where the Nusselt number $N = \frac{1}{\Gamma} \int_0^\Gamma dx\, Q_z$ is the total vertical heat current through the fluid layer, $\int_0^\Gamma dx\, Q_z$, reduced by the conductive part, $\int_0^\Gamma dx\, Q_{cond} = \Gamma$. In our scaling, $Q_{cond} = 1$.

The stability properties of the conductive state against infinitesimal convective perturbations [67–70] are summarized for ethanol-water parameters, $L = 0.01$ and $\sigma = 10$, in [46, Fig. 1] as a function of $\psi$ for the experimentally-accessible $\psi$-range. We use the scaled Rayleigh number $r = \frac{R}{R_c^0}$ that is reduced by the critical Rayleigh number $R_c^0$ for onset of convection in a pure fluid with the critical wave number $k_c^0$. The analytical values are $R_c^0 = 1707.762$ and $k_c^0 = 3.11632$. In the finite-differences approximation of the field equations of our MAC algorithm, however, the threshold for onset of pure fluid convection [71] lies at $R_c^0 = 1686(\pm 0.2\%)$ for $\Delta x = \Delta z = \frac{1}{20}$ with which we scale our Rayleigh numbers here.

## III. LOCALIZED TRAVELING WAVE STATES

In this section, we discuss common properties of LTW's that we have obtained for a weaker Soret coupling of $\psi = -0.08$ and a stronger one of $\psi = -0.25$ as stable solutions of the 2d hydrodynamic field equations in a vertical cross-section perpendicular to the roll axes. In fact, experiments so far seem to suggest that the proximity of walls in narrow convection channels which orient the axes of the convection rolls perpendicular to the walls and parallel to each other – thus enforcing a configuration we are simulating – is necessary to stabilize spatially-localized traveling convection rolls [24].

In Fig. 1, we show in the bifurcation diagrams of frequency and maximum flow velocity versus Rayleigh number the location of our LTW attractors by open circles in comparison with the attractors of extended TW states (full circles) and of extended SOC states (full squares). For better comparison, the reduced Rayleigh numbers are shown relative to the respective oscillatory convection thresholds $r_{osc}$. For the stronger Soret coupling $\psi = -0.25$, the LTW's are subcritical with respect to $r_{osc}$, and they compete with extended TW states and the stable quiescent conductive state. For the weaker coupling $\psi = -0.08$, LTW's lie above $r_{osc}$ in a regime of Rayleigh numbers where the conductive state is only convectively unstable [68,70] but not yet absolutely unstable. Here, the competing extended state attractor is not TW convection but SOC convection — for $\psi = -0.08$, the band of stable LTW's lies above the transition $r^*$ where stable extended nonlinear TW's cease to exist and SOC states become stable [46].

Despite these differences — the states at $\psi = -0.08$ and $-0.25$ are discussed in detail in sec. IV and V, respectively — the LTW's display similar physical properties to be discussed in this section, e.g., a nontrivial spatio-temporal symmetry. We compare the field structure of our LTW's in the $x - z$ plane perpendicular to the roll axes with each other and with extended TW states. Furthermore, we elucidate the effect of the large-scale, current-induced concentration redistribution on the buoyancy profile of the LTW and its consequences.

### A. Structure

Here we first briefly review and then expand our results [42,26,27] on the structure of LTW's. While they differ in spatial extent and somewhat less in frequency, group velocity, and convection intensity, they exhibit similar structure and behavior.

#### 1. Fields

If one considers LTW's to consist of three characteristic parts — a leading front, a central part, and a trailing front, relative to the propagation direction of the TW phase — then one can roughly say that all our LTW's differ only in the extent of their central part. Structural properties at the leading front seem to be universal. The trailing fronts of the LTW states also resemble each other, but the field structure there is distinctly different from the leading front. These features can be seen in Fig. 2, where narrow LTW's at $\psi = -0.25$ and $\psi = -0.08$ with different widths are shown next to each other. In fact, the structural properties at the leading and trailing fronts, respectively, of the two different LTW's agree with each other even in detail. This is documented in Fig. 2, where we compare (a) the shape of the wavelength variation $\lambda(x)$, (b) the vertical velocity field $w(x, z = 0.5)$, (c) the streamlines of the large-scale mean concentration current, and (d) the lateral profiles of the time averaged convective temperature $\langle\theta\rangle$, concentration $\langle c\rangle$, and buoyancy force $\langle b\rangle = \langle c + \theta\rangle$, to be discussed further below. For a more detailed presentation of the structure of the various fields, see [42,26,27].

The topview shadowgraph intensity distributions

$$I(x) = A \int_0^1 dz\, \partial_x^2 (\delta C + b\,\delta T) \qquad (3.1)$$

of LTW's obtained in recent experiments [9,10,15] show structural properties [72] that agree nicely with our predictions for the LTW fields $\delta C$ and $\delta T$ entering into (3.1). $A$ is a constant and $b = -0.919$ with our parameters for a real mixture [73,62].

After transients have died out, all our LTW's drift with a finite but small group velocity $v_g$ forwards, i.e., into the propagation direction of the phase of its TW constituents, while the width $\ell$ remains constant. In the coordinate system $\breve{\Sigma}$ comoving with the center of mass velocity $v_g$, the LTW fields are periodic in time:

$$\delta \breve{F}(\breve{x}, z; t) = \delta \breve{F}(\breve{x}, z; t + \breve{\tau}), \qquad (3.2)$$

with $\breve{\tau}$ being the oscillation period of the LTW in its rest frame $\breve{\Sigma}$. Remarkably enough, our LTW states display an additional nontrivial spatio-temporal symmetry

$$\delta \breve{F}(\breve{x}, z; t) = \pm \delta \breve{F}\left(\breve{x}, 1-z; t + \frac{\breve{\tau}}{2}\right) \qquad (3.3)$$

with $+$ for $u$, $p$ and $-$ for $w$, $\delta T$, $\delta C$, and $B$. This symmetry (3.3) of LTW's is the analog of the symmetry [46, eq.(3.16)] of extended TW's.

### 2. LTW versus TW

The fields in the center of LTW states have a similar structure to that in extended TW states. This is demonstrated in Fig. 3, where we compare a LTW state and a TW state for the same parameters $L = 0.01$, $\sigma = 10$, $\psi = -0.25$, $r = 1.246$. The TW state is the one of [46, Fig. 3b], here however presented as moving to the left like the LTW. The LTW is the one of Fig. 2 (left column), however, presented at a different time, so that the node positions in Figs. 2 and 3 are different. The LTW frequency $\omega_{LTW} \approx 4.3$ is much bigger than the TW frequency $\omega_{TW} \approx 2.2$. Furthermore, the LTW wavelength (left column of Fig. 2a) varies laterally [42,9,17,10,15] and is about $\lambda \approx 1.8$ in the center part near $x = 0$ of the LTW $A_1$, as compared to $\lambda = 2$ in the extended state. Nevertheless, the LTW structure in the center is very similar to that of the TW. Compare, e.g., the colour-coded concentration fields with their plume structures and trapezoidal lateral profiles, or the sideview shadowgraph intensity distributions in the respective rows of Fig. 3. Since the phase velocity of the LTW is bigger than of the TW, the concentration contrast between adjacent rolls is significantly higher in the former than in the latter state. The wavelength [72] of the LTW increases towards the leading front (left column of Fig. 2a) and with it the local phase velocity. The increase of $v_p$ also causes, for reasons explained in [46], the concentration contrast between adjacent rolls to increase – see the lateral LTW concentration profile in Fig. 3. Thus the mean lateral concentration current is also increased.

### B. Buoyancy balance and the large-scale current-induced concentration redistribution

Analysing the balance of buoyancy forces that drive convection is central to understanding the dynamics of LTW's. So, in this section we demonstrate how the distribution of the mean concentration (averaged over one oscillation period of the LTW) that is associated with a large-scale, closed, mean concentration current circulating over the whole LTW state influences the local effective driving force for convection.

### 1. Time averages

We average over one oscillation period $\tau = \frac{2\pi}{\omega}$ of the LTW. Since the spatial variation of the wave number, together with the finite, albeit small group velocity, causes a small ambiguity in the definition of the frequency in the laboratory system, we always define $\omega$ in the center of the LTW state. The average

$$\langle \delta F \rangle = \frac{1}{\tau} \int_0^\tau \delta F(x, z; t') \, dt' \qquad (3.4)$$

acts like a low pass filter that only the slowly-varying parts of the fields can pass. The fields (3.4) are not constant in time. But, since $v_g \ll v_p$, they are very similar to the properly averaged fields in the system $\breve{\Sigma}$ comoving with the LTW's group velocity

$$\left\langle \delta \breve{F} \right\rangle (\breve{x}, z) = \frac{1}{\breve{\tau}} \int_0^{\breve{\tau}} \delta \breve{F}(\breve{x}, z; t') \, dt'. \qquad (3.5)$$

Because of (3.3), these averaged fields are stationary and show the symmetry

$$\left\langle \delta \breve{F} \right\rangle (\breve{x}, z) = \pm \left\langle \delta \breve{F} \right\rangle (\breve{x}, 1-z) \qquad (3.6)$$

with $+$ for $u$, $p$ and $-$ for $w$, $\delta T$, $\delta C$, and $B$.

In the center part of the LTW, as in an extended TW, convection reduces the time averaged vertical temperature and the concentration gradient in the bulk of the fluid layer. Fig. 2d presents evidence for this convectively induced flattening of the vertical field profiles in the pulse centers: in the lower warm and alcohol-poor region of the layer, say, at $z = 0.25$ one has $\langle \theta \rangle < 0$ and $\langle c \rangle > 0$, while in the upper cold and alcohol-rich region of the layer, the reverse holds, according to (3.6). Fig. 3, 3rd row, shows the convective homogenization of the mean concentration by the green color in the center part of the LTW as compared to the strong vertical gradient in the surrounding quiescent conductive regions.

### 2. Buoyancy force profiles

The above described convection-induced vertical concentration redistribution enhances the mean buoyancy force $\langle b \rangle = \langle c + \theta \rangle$ (cf. Fig. 2d) relative to that in the conductive region, while the convective temperature equilibration in the bulk of the layer has the opposite effect of weakening $\langle b \rangle$. Note that a positive (negative) $\langle b \rangle$ in the lower (upper) half of the fluid layer implies an enhancement of the force $\langle b \rangle \mathbf{e_z}$ in upwards (downwards) direction.

It is obvious from Fig. 2d that the convection-induced mean buoyancy profile $\langle b \rangle$ of the LTW is dominated by the concentration contribution. Furthermore, since the temperature pulse is significantly smaller than the concentration pulse (cf. Fig. 2d) one observes: (i) far outside the center part of the LTW, where $\langle \theta \rangle$ vanishes, it is the concentration distribution $\langle c \rangle$ alone that determines the buoyancy $\langle b \rangle$, and (ii) the combination $\langle c + \theta \rangle$ is largest right under the two fronts (cf. the extrema in Fig. 2d). This increase of the local buoyancy force $\langle b \rangle$ right under leading and trailing fronts stabilizes the LTW at both fronts against an invasion of the surrounding conductive state. On the other hand, ahead of the leading front, the concentration distribution is such as to weaken the buoyancy – the negative dip in $\langle c \rangle$ and $\langle b \rangle$ located at $x \approx -4.5$ in the left LTW of Fig. 2d implies in the lower half of the fluid layer a depression of the concentration and of the *upwards* buoyancy force relative to the conductive state. By symmetry, $\langle c \rangle$ and $\langle b \rangle$ are positive in the upper half, thus depressing the *downwards* buoyancy. The concentration depression in the lower half of the layer ahead of the front at $x \approx -4.5$ can also be seen in the snapshot of the LTW concentration field in the 4th row of Fig. 3a. The large, red, upwards-bending bulge reflects a concentration-depressed region that contains less alcohol than in the conductive state. Half a period later, one would observe a correspondingly-shaped blue bulge at $x \approx -4.5$ in the upper half of the layer with alcohol concentration above the conductive reference level. This peculiar concentration distribution ahead of the leading LTW front acts like a "barrier". By reducing the local buoyancy force there, it stabilizes the conductive region ahead of it against the invasion of convection located behind it.

To conclude, the vertical profile of the time-averaged concentration "barrier" ahead of the leading front shows a surplus, $\langle c \rangle > 0$, in the upper half of the layer and symmetrically a depletion, $\langle c \rangle < 0$, in the lower half relative to the conductive concentration distribution. So in the "barrier" the vertical concentration gradient is enhanced.

### 3. Mean concentration current

The presence of this concentration "barrier" ahead of the leading front can be related to the existence of a strong mean circulating concentration current

$$\langle \mathbf{J} \rangle = \langle \mathbf{u}\, \delta C \rangle - L \boldsymbol{\nabla}\, \langle \delta C - \psi\, \delta T \rangle \; . \qquad (3.7)$$

Its streamlines are shown in Fig. 2c and in the third row of Fig. 3a. The origin of this current is the phase shift between the concentration field and the velocity field in the center part of the LTW. This shift leads, as in extended TW's [46], to a strong time-averaged concentration current in the center of the LTW. In the upper half of the layer, $\langle \mathbf{J} \rangle$ flows parallel to the phase velocity of the LTW, i.e., here to the left, and vice versa in the lower half of the layer. So $\langle \mathbf{J} \rangle$ has nearly horizontal streamlines in the central part of the LTW. On the other hand, there is almost no concentration current in the conductive area surrounding the LTW. Hence, the two currents that flow oppositely to each other in the center part of the LTW bend over vertically, thus forming a large-scale, closed circulation loop (dashed lines in Fig. 2c) extending over the whole LTW. The horizontal currents in the center part are convective, while the vertical currents are partly diffusive.

In addition to the large-scale concentration current that circulates counterclockwise along the dashed streamlines in Fig. 2c, there are two small, secondary loops with clockwise circulation under the leading front. They are caused by the strong convective reduction (increase) of the mean concentration $\langle \delta C \rangle$ in the center part of the LTW in the upper (lower) half of the fluid layer relative to the surrounding conductive state. Near the plates, where diffusion dominates, the resulting lateral concentration gradients drive a lateral diffusive current

$$\langle \mathbf{J} \rangle = -L\, \partial_x\, \langle \delta C \rangle\, \mathbf{e}_x \qquad \text{at } z = 0, 1 \, , \qquad (3.8)$$

which transports concentration out of the conductive region into the LTW at the upper plate and vice versa at the lower plate. This increases slightly the primary concentration circulation at the trailing front; see the small dents in the outer concentration streamlines. But at the leading front this diffusive effect counteracts the primary concentration current and generates the two small secondary countercirculating current loops (solid lines in Fig. 2c) under the leading front. Spurs of very low intensity of the primary, left-turning concentration circulation, that reach into the conductive region, are not resolved in our plot.

### 4. Concentration current and drift velocity

All in all, the time-averaged concentration current sustains a small concentration surplus in the upper half of the layer ahead of the leading front and a small concentration deficiency in the lower half of the layer, as compared to the conductive state, which reduces the buoyancy force there. This current-induced "barrier" seems to be the reason why the LTW drifts with a much smaller group velocity $v_g$ than the phase velocity of the rolls. This explanation is supported by the experimental and numerical fact that $v_g$ increases as the phase velocity, i.e., the LTW frequency decreases with growing $r$: As $\omega$ decreases, so does the mean concentration current, cf. [46, Fig. 8]. Thus the concentration redistribution forming the "barrier" and its "braking" effect becomes weaker.

Also, the experimentally-observed [11,10,12,14,15] reduction of $v_g$ even below zero for smaller $r$, i.e., larger $\omega$, fits into this picture: For larger $\omega$, the concentration-current-induced, buoyancy-reducing vertical concentration gradient ahead of the leading front becomes so strong that convection cannot penetrate into the conductive state at the leading front and is even — for sufficiently large $\omega$ (small $r$) — pushed back by the expanding conductive state, so that the LTW drifts backwards (see also the discussion in sec. IV.A).

### 5. Further mean transport

(a) *Mean heat transport*— The phase difference between temperature wave and velocity wave causes a time-averaged *lateral* heat current $\langle Q_x \rangle$ in the center part of the LTW as discussed in section III.E for extended TW's. But, since there is also a large mean *vertical* heat current $\langle Q_z \rangle$, heat is transported mainly upwards. The lateral heat current $\langle Q_x \rangle$ only leads to a lateral bending of the streamlines of $\langle \mathbf{Q} \rangle$ and also of the mean convective heat current $\langle \mathbf{Q} \rangle - \mathbf{Q}_{cond}$ in the center part of the LTW. All in all, the time-averaged heat current $\langle \mathbf{Q} \rangle$ has, with its open, nearly vertical streamlines, a strongly different structure than the concentration current $\langle \mathbf{J} \rangle$. In particular, there is no *circulation* of heat.

(b) *Mean flow?*— While extended TW's do generate a small lateral mean flow (section III.E), its LTW analog would drive a kind of circulating Poiseuille flow in the conductive region surrounding the convective part of the LTW state in annular geometry. The symmetry (3.6) only yields the information that, in the system $\breve{\Sigma}$ comoving with the group velocity $v_g$ of the LTW,

$$\langle \breve{u} \rangle (\breve{x}, z) = + \langle \breve{u} \rangle (\breve{x}, 1-z) \qquad (3.9)$$

and

$$\langle \breve{w} \rangle (\breve{x}, z) = - \langle \breve{w} \rangle (\breve{x}, 1-z) . \qquad (3.10)$$

For our calculation accuracy, the MAC algorithm shows an extremely small lateral mean flow of the order of $10^{-5}$; remember that the mean flow in extended TW states is of the order of $10^{-3}$ [51,46,74]. We cannot exclude a physical origin for the mean flow in the LTW's, but we suppose that it is a numerical artifact, which has its source in the iterative adjustment of the velocity fields and the pressure to each other. In this procedure, incompressibility of the fluid is guaranteed only within a given accuracy limit.

### C. Other theoretical work

Recently, pulse solutions [28,40,31,32,37,41,38,1] have been found in amplitude equation models containing unsystematically some fifth-order terms. However, these models should not be expected to yield a full, realistic description of the structure and dynamics of LTW fields in binary mixtures, since one common amplitude $A(x, t)$ for all fields multiplying a harmonic wave $e^{i(k_c x - \omega_c t)}$ cannot represent the fields appropriately — TW's and LTW's, being strongly nonlinear states, differ significantly from linear convective plane-wave perturbations. Furthermore, the concentration field, with its nonharmonic structure that differs dramatically from the temperature and velocity fields, controls the buoyancy dynamics described in the previous section. Thus, approaches using single-mode amplitude models fail to incorporate the LTW dynamics related to the large-scale concentration redistribution. Indeed, this seems to be one reason for the fact that the pulses of simple amplitude equation models have a fast group velocity, comparable in magnitude with their phase velocity. Such equations, when enlarged [44,45] to contain a mean concentration mode in an idealized way, i.e., under FSP conditions, show a concentration redistribution effect and thus a slowing down of the group velocity. This seems to be in line with our earlier explanation [42] that the concentration redistribution in real LTW's, described above, leads to a stabilization of the conductive (convective) region at the leading (trailing) edge against penetration of convection (conduction). The propagation of the two fronts separating the two regions is thereby hindered. Derivation of a model for realistic, impermeable boundary conditions that would ensure an appropriate concentration balance without allowing leakage through the plates and more quantitative tests of its results would be very useful to assess its validity and predictive power.

Considerations of concentration currents [75,2] and speculations on a concentration redistribution [50,76] in connection with lateral concentration currents [77,50] have been presented earlier.

## IV. NARROW LTW PULSES AT $\psi = -0.08$

In this section, we discuss bifurcation behavior, generation, stability, and decay of LTW's for $\psi = -0.08$. We start with this $\psi$ because the LTW properties are simpler here than for $\psi = -0.25$. In experiments [6,13,10,15], only LTW states of selected small width, like the state $X$ on the right side of Fig. 2, have been found for $\psi$ around $-0.08$. We get the same result with our simulations.

### A. Bifurcation properties compared to experiments

For the Rayleigh numbers $r_X = 1.104$, $r_Y = 1.106$, $r_Z = 1.109$ above $r_{osc}$ where we performed calculations, we found only narrow LTW pulses (Fig. 4 and Tab. I). One of them, namely, $X$ has been described in [42,26]. They compete with stable extended SOC states — here

$r^*$ lies below $r_{osc}$, cf. [46, Fig. 9]. We have not systematically determined the basin of attraction of these different stable states by varying initial conditions. For $r = 1.099$ and $1.102$ below $r_X$ and for $r = 1.111$ above $r_Z$, LTW's were unstable in a characteristic manner. So we find stable LTW states in an $r$ interval which is larger than 0.005 but smaller than 0.009. In experiments, this $r$ band is about 0.02, i.e., at least twice as big. Our LTW band lies clearly above $r_{osc}$, cf. Fig. 1. This also holds for the experimental LTW's [10] for $\psi = -0.072$. Niemela et al. [6] have observed in their experiments for $\psi = -0.08$ and Kolodner [10] for $\psi = -0.102$ that the LTW band also reaches somewhat into the "subcritical" region below $r_{osc}$. For $\psi = -0.123$, the experimental LTW band [10] lies substantially below $r_{osc}$ and reaches only somewhat into the "supercritical" region above $r_{osc}$. The behavior at $\psi = -0.127$ [15] is nearly identical to that at $\psi = -0.123$.

The maximal flow amplitude $w_{\max}$ of our LTW's (Fig. 4b) increases by about 20% in the LTW band. At the lower band end, $w_{\max}$ reaches about 75% of the competing SOC flow amplitude and at the upper end about 82%. The experiments for $\psi = -0.123$ also show an increase with $r$. But the amplitude variation in the experiments seems to be much stronger than in our calculations since the maximal amplitude of the experimental shadowgraph signal doubles [10, Fig. 17a].

We define the width $\ell$ of our pulses by the full width at half maximum (FWHM) of the intensity envelope of the vertical velocity field $w$ at midheight of the fluid layer. The similarly determined width of the shadowgraph intensity envelope $I(x)$ obtained from (3.1) is for our states with $\psi = -0.08$ by about 0.3 slightly smaller than $\ell$. The main reason for this difference is the second derivative of the concentration field that enters into $I(x)$: whereas the LTW of the concentration field is broader than the LTW of the temperature field which itself is broader than that of the velocity field (cf. the profiles of $w$, $\langle\theta\rangle$, and $\langle c\rangle$ in Fig. 2 or in [42]; for more details see [27]) — this is not the case for the second derivatives, since the concentration field is much smoother at the fronts than in the center of the LTW. Note that an experimentally determined LTW width $\ell_{exp}$ measures the FWHM of the shadowgraph intensity $I(x)$, where a low pass filter is sometimes used; this procedure reduces the influence of the concentration field [73]. As an aside we mention that the width of the pure temperature contribution to $I(x)$ is here and also for the LTW states with $\psi = -0.25$ nearly identical to $\ell$. Fig. 4c shows that $\ell$ is nearly constant, exhibiting a small increase $\approx 5\%$ with $r$. The LTW's of Niemela et al. [6] for $\psi = -0.08$ also show a constant width within the experimental accuracy. An inspection of [10, Fig. 17b] and [15, Fig's. 5 and 32] also seems to suggest a constant or slightly increasing width with $r$ for Kolodner's experiments at $\psi = -0.123$ and $-0.127$. The widths of these experimental pulses seem to be of the same magnitude as our numerical pulses at $\psi = -0.08$. But the $r$-band of experimentally stable narrow LTW's at $\psi = -0.127$ [15, Fig's. 5 and 32] is considerably wider than of the numerical pulses at $\psi = -0.08$.

The group velocity $v_g$ (Fig. 4d) of the numerical pulses is small and grows sublinearly. This agrees with experiments [10]. Our group velocities are positive, i.e., parallel to the phase velocity, as in the experimental mixtures with $\psi = -0.072$ and $\psi = -0.101$. For $\psi = -0.123$, Kolodner reports a small $r$ region of negative group velocity. The numerical $v_g$ at $\psi = -0.08$ lies slightly above the experimental one at $\psi = -0.072$ [10, Fig. 11]. Indeed, for the smaller $r$ values, the experimental $v_g$ reaches down to zero. But, all in all, experiment and simulation show group velocities which have the same order of magnitude.

Kolodner [11,10] explains the pinning of LTW pulses in earlier experiments, i.e., the reduction of $v_g$ to zero, with too large spatial variations of the local Rayleigh number in the experimental cell. But also those convection cells in which pulses do travel still have small inhomogeneities [10] which might explain part of the differences in the group velocity between simulation and experiment. Furthermore, experiments in narrow annular channels observe 3d LTW convection which is influenced by the radial sidewalls, whereas the calculation simulates truly 2d convection.

Finally, there are also differences in the experimental and numerical Lewis and Prandtl numbers, e.g., in experiments, $L$ is mostly smaller than 0.01. It is tempting to speculate about the consequences of the smaller $L$ in experiments: (i) It will cause a larger concentration contrast in the TW pattern, thus a larger oscillation frequency, thereby a larger mean horizontal concentration current, and a larger concentration redistribution. (ii) Diffusion being reduced, the concentration "barrier" ahead of the pulse described in Sec. III.B.3 will tend to be stronger. While both effects have the tendency to slow down the pulse, we do not know whether their magnitude can explain quantitatively the difference between the faster numerical drift and the slower experimental one.

The LTW frequency $\breve{\omega}$ (Fig. 4a) in the comoving frame $\breve{\Sigma}$ was determined unambiguously from half the frequency of the Nusselt number. The latter oscillates slightly because convection rolls are created under the trailing front and annihilated under the leading front. The LTW frequency decreases strongly with growing $r$. Such an $r$ dependence has also been observed [17] for LTW states resting in the middle of small rectangular cells and very recently [15, Fig. 1b] for drifting pulses in an annular channel at $\psi = -0.127$. Note that the increase of $v_g$ and the reduction of $\breve{\omega}$ with $r$ lead to a yet stronger decrease of the ratio of phase velocity to group velocity from 13.5 for state $X$ to 6.5 for state $Z$.

## B. Why LTW's are stable above $r_{osc}$

Our LTW's for $\psi = -0.08$ and many experimental LTW's exist above $r_{osc}$ in the convectively unstable regime [68,70]. So spatially confined TW convection stably coexists with the surrounding conductive state, which is unstable at $r > r_{osc}$ against *extended* TW perturbations. To explain this, it has been proposed [16,42,26], experimentally seen [7,8] and analysed [21,22] that with laterally periodic boundaries, initially *localized* disturbances of the conductive state, growing as weakly nonlinear structures, move with the fast linear group velocity into the LTW. They are absorbed in the LTW state and so do not have enough time to grow to an extended state. An elegant way to reduce effectively the time/space that is available for the disturbances to grow before they are swept into and absorbed by an LTW is to prepare [10] more than one LTW in an annular container. Such a multi LTW state is stable to higher r values than a single LTW state. With the smaller mean travel time/distance of disturbances before absorption, they do not reach a dangerous size. See also the experiments and the discussion of Kaplan and Steinberg [25] of the question of how LTW's grow unstable at the upper end of the stable $r$ band for various lengths of (rectangular) cells.

LTW's are stable against weak disturbances, and only stronger disturbances destroy them [21]. So the reasons that disturbances do not have enough time to grow are (i) the conductive state is only convectively but not absolutely unstable, (ii) the group velocity of linear disturbances of the conductive state is much larger than that of the strongly nonlinear LTW, and (iii) the absorption capabilities of the latter. For further discussions, see Kolodner and Glazier [21] and Steinberg and Kaplan [17,25]. But these arguments do not explain the fixed width of the LTW pulses, i.e., why and how the leading and trailing fronts of the LTW are coupled to have the same velocity.

## C. Generation of LTW's

To generate the first LTW $X$, we "filled" half of our system of length $\Gamma = 20$ with an extended SOC state and the other half with the conductive state. These initial conditions evolved at $r_X$ into the LTW state $X$. To show that this state is uniquely selected, as we expected from experiments [6], we proceeded from state $X$ as follows: We increased the drive to $r = 1.145$, i.e., clearly above the stable LTW band. There, the phase velocity strongly decreased, and the width of the convective region rapidly increased. After 50 diffusion times, it was about three times as big as at the beginning and nearly filled the entire system. The frequency was about $\frac{1}{8}\omega_H$ and the wavelength about 2 in the convective area. This transient solution — which we checked in a different run at the fixed $r = 1.145$ to evolve into a transient slow TW relaxing towards the stable extended SOC state — was then used as initial configuration for reducing the drive down to $r_X$. With this reduction, the LTW state $X$ developed again. Since this time, however, the initial condition was completely different, we conclude that there is only one stable LTW available at $r_X$, namely the state $X$.

The transition from the broad and slowly-traveling transient to state $X$, as shown in Fig. 5, is typical. After reduction of $r$ to $r_X$, the velocity, temperature, and Nusselt number adapted within 1 or 2 diffusion times to the new $r$ value. We determined an intensity center $X(t)$ of the velocity field, not to be confused with the state $X$, by

$$X(t) = \frac{\int dx\, x\, w^2(x, z = 0.5; t)}{\int dx\, w^2(x, z = 0.5; t)} \ . \qquad (4.1)$$

It travels only slightly during this time. The standard deviation $\Delta X(t)$ defined via the second moment of $w^2$,

$$(\Delta X(t))^2 = \frac{\int dx\, x^2\, w^2(x, z = 0.5; t)}{\int dx\, w^2(x, z = 0.5; t)} - X^2(t) \ , \qquad (4.2)$$

decreases by about 15% during this short time (Fig. 5d). During the next 100 diffusion times, $X(t)$ traveled with a group velocity of about 0.07. Since the leading front moved with a velocity of about 0.05 and the trailing front with about 0.09, the width of the LTW state decreased by about 10 to 15%. The phase velocity (see $\breve{\omega}$ in Fig. 5a) got with about 0.14 twice as large as the group velocity (Fig. 5c), while the wavelength remained about 2. During the next 100 diffusion times, the width halved and $N - 1$ (Fig. 5b) decreased by a factor of three, while the phase velocity more than doubled. The group velocity increased further to a maximum value of about 0.13 at $t \approx 200$. By then, the transient LTW had nearly reached its end *structure*. The higher phase velocity has lead to a stronger concentration contrast between adjacent rolls. The wavelength $\lambda$ now varies already strongly in the convective region. The maximal upward flow amplitude $w_{\max}$ has decreased by about 15%.

In the time interval from 200 to 250, the end state $X$ was reached. During this time, the width of the state decreased minimally. But the phase velocity doubled, the group velocity nearly halved, $N - 1$ further decreased by about 25% and $w_{\max}$ by 10%. Connected with the increase of the phase velocity, the concentration contrast between adjacent rolls increased again. The transition to the final state happened in a characteristic manner. The change of the group velocity happened between $t = 200$ and $t = 220$. Simultaneously, $N(t) - 1$ and the width of the state showed a slight "undershooting" under the value of the stationary LTW, combined with a sharp decrease of the group velocity. It is not clear whether this "undershooting" is a hint of an oscillatory relaxation into the stable LTW end state, since, if there are indeed such oscillations, then they lie within the order of magnitude of the oscillations generated by the propagation of the rolls under the front. In any case, we did not observe such an "undershooting" for $\psi = -0.25$ (cf. sec. V.D.). Since the velocity of the *leading* front stayed nearly constant during the entire process, the behavior of the width and also the "undershooting" in the width of the state by about 5% reflects the time behavior of the *trailing* front.

At the upper end $r_Z$ of the LTW band, we observed the selection of the narrow LTW state $Z$. Starting with a broad LTW state, we saw qualitatively the same behavior there as we have just described for the state $X$. At $r_Y$, we investigated the transition behavior in the LTW band. Starting from the state $X$, the system reached the stable end state $Y$ about 50 diffusion times after the increase of the drive from $r_X$ to $r_Y$, exhibiting a simple relaxation behavior.

### D. Decay of LTW's

The fate and evolution of LTW's differs significantly after crossing the lower or upper stability boundary of the stable LTW band. We first consider the decay of an LTW pulse at the lower end of the LTW band. Starting from the state $X$, we reduced the drive by about 0.002 to $r = 1.102$, where the LTW state is unstable and decays towards the conductive state. But still being above $r_{osc}$, the conductive state is also unstable. (As an aside, we mention that, after reduction to $r = 1.084 < r_{osc}$, the system ended in the then stable conductive state although there is also a stable extended TW state – cf. [42, Fig. 1].) Let us now consider the behavior in more detail. The LTW retains its width up to the time $t \approx 60$, while the strength of convection decreases by about 30%. The group velocity decreases to about 0.03, which we would extrapolate from Fig. 4d for a LTW at this $r$ value. Over the next 20 diffusion times, convection nearly dies out. But with the conductive state being unstable, convection starts to grow again, predominantly at the leading front, and there is an interplay between decay of the LTW and growing extended convection. Now the concentration field no longer has the trapezoidal structure, but is nearly harmonic, like the temperature and velocity fields. At $t = 90$, an extended transient TW has developed with strong amplitude modulation. The strength of the modulation decays fast – for a more detailed description, see [27, chapter 8.2.2.2]. At $t \approx 150$, the amplitude of the transient TW is still very small and grows exponentially. At $t \approx 170$, the relaxation into the "strongly nonlinear" extended SOC state begins. Together with the increase of the amplitude, the phase velocity decreases and decays in the long time limit to zero. So, when crossing the lower band limit of stable LTW's, the convection intensity first decreases and then grows again.

At the upper end of the LTW band, the dynamics of unstable LTW's is completely different and similar to what has been seen by Kolodner [10]. There, the LTW expands, and finally its fronts touch one another. Then, the gap between the two fronts closes, and a slow, spatially modulated, extended transient TW develops which then relaxes into the stable SOC state. For longer convection cells, one would expect that the "filling" of the conductive region with convection could also be determined by growing disturbances of the conductive state. See the experimental work of Kolodner and Glazier [21,22] and Kaplan and Steinberg [17,25].

### V. LTW'S AT $\psi = -0.25$

For more negative Soret coupling, at $\psi = -0.25$, LTW's show richer and more complex bifurcation properties than at $\psi = -0.08$. The situation is indeed more complicated than we expected from the early experiments [4,5,9] and also seems to be more intricate than the one reported in recent experiments [14,15]. We found on the one hand stability of different LTW states with different widths in a narrow interval of Rayleigh numbers and on the other hand below and above it uniquely selected LTW states with a uniquely selected width, indicating monostability.

### A. Properties

Since the determination of complete LTW bifurcation diagrams clearly exceeds our computational resources, we discuss a skeleton (Fig. 6) based on 12 stable LTW states $A_1, A_2, \ldots, F_1, F_2$ that we have generated at four different Rayleigh numbers. All of them lie well below $r_{osc}$, as expected from experiments. Their characteristic properties are listed in Tab. II. States $A_1, B_1, C_1$ have been described in [42,26]. These solutions of the field equations were obtained with laterally periodic boundary conditions and periodicity lengths $\Gamma = 20$ (circles in Fig. 6) or $\Gamma = 40$ (triangles in Fig. 6). Some of them are identical, for all practical purposes, e.g., state $D_1$ obtained with $\Gamma = 20$ is the same as state $D_2$ obtained with $\Gamma = 40$. The index either identifies $\Gamma$ as in the case of $D_1, D_2$ and/or it identifies different histories that lead to the respective final states, as explained in the text. For example, states $E_1 = E_2$, both in $\Gamma = 20$ systems, have different histories. Now we list and discuss the problems and questions (a) - (g) that we have addressed together with our findings.

(a) *Multistability.* — Bensimon et al. [4,5] and Surko et al. [9] reported experimental LTW states within an interval $\Delta r \approx 0.02$ that were stable with continuously different widths. We have shown [42,26] that at least for one Rayleigh number, $r_A = 1.246$, there are stable states $A_1, B_1$ in a $\Gamma = 20$ system and $C_1$ in a $\Gamma = 40$ system having different widths. Strength of convection $w_{max}$, frequency $\breve{\omega}$, and group velocity $v_g$ are essentially the same for these states $A_1, B_1, C_1$. For more details on these different states at $r_A$ and our conclusion that there is in fact bistability at $r_A$, cf. point (f) and (g) below.

(b) *Finite r-band of stable LTW states.* — Our states at $r_D = 1.241$ and $r_E = 1.244$ below $r_A$ and at $r_F = 1.248$ above $r_A$ show that there is a finite $r$-band in which LTW's exist. Within this band, the frequency $\breve{\omega}$ does not vary much, while $w_{max}$ (Fig. 6a) and $v_g$ (Fig. 6c) increases slightly with $r$. For all states, $v_g$ is small and positive, i.e., parallel to the phase velocity. Their width is discussed further below. While we did not determine the lower existence limit of the $r$-band, we can say that it lies between $r_D = 1.241$ and $1.2155$. At the latter Rayleigh number, which is just above $r_{TW}^s$, a broad, transient LTW shrank to a narrow transient LTW which then decayed into the conductive state. Concerning the upper band limit, we can say that, in a $\Gamma = 20$ system, it lies between $r_F = 1.248$ and $1.251$: Starting with state $B_1$ at $r_B = 1.246$ and increasing $r$ to $1.251$, convection spread in both directions into the conductive region. In a $\Gamma = 40$ system, on the other hand, some test calculations described in (e) below give a hint that the upper band limit for stable LTW's there lies at a smaller $r$, somewhere between $r_A = 1.246$ and $r_F = 1.248$.

(c) *Irregular behavior.* — When leading and trailing fronts of the above described expanding LTW transient in a system length $\Gamma = 20$ at $r = 1.251$ touched each other, they interacted strongly. In this case, width, shape, and Nusselt number showed complicated, irregular behavior [27] for several hundred thermal diffusion times. We have hints that this aperiodicity is not transient but genuine, long-time behavior: (i) Upon reducing $r$ to $r_A = 1.246$, the system did not show a transition to a regular LTW state but continued to behave erratically, so we stopped the simulation. (ii) At $r = 1.256$ we also observed such an irregular behavior for more than $1000$ diffusion times before we stopped the calculation. For $r = 1.266$, this front interaction period lasted only about $200$ diffusion times. Then, convection invaded the entire space, and the system relaxed into a stable extended TW with $10$ roll pairs.

For $\psi = -0.08$, we did not find erratic behavior triggered by the interaction of the LTW fronts. This novel phenomenon does not seem to have been reported before. For instance, the erratic behaviour discussed in [15, Sec. VIII] seems to be a bulk phenomenon that is not related to an interaction of leading and trailing front that touch each other in our periodic system.

(d) *Narrow LTW states below $r_A$.* — The narrow LTW $D_1$ evolved in a $\Gamma = 20$ system out of $B_1$ after instantaneous reduction of $r$ from $r_A$ to $r_D$, while $D_2$ evolved in a $\Gamma = 40$ system out of a very broad transient LTW that filled half of the entire space. The narrow final states being practically identical strongly suggests monostability or a unique selection mechanism at $r_D$ [27]. Also, the LTW's $E_1 = E_2$ at $r_E$ seem to be uniquely selected: $E_1$ resulted from $B_1$ after instantaneously decreasing $r$ from $r_A$ to $r_E$, while $E_2$ developed out of $B_1$ as $r$ was very smoothly ramped down from $r_A$ to $r_E$. Note that $B_1$ is a broad state, while $D_1 = D_2$ and $E_1 = E_2$ are narrow states.

(e) *Broad LTW states above $r_A$.* — We also found monostability or unique selection above $r_A$ in a system of length $\Gamma = 20$ — the two states $F_1 = F_2$ at $r_F = 1.248$ have completely different histories [27]. $F_1$ developed out of $B_1$, which has a large width comparable with $F_1$, whereas $F_2$ evolved out of $E_1$, which has a much smaller width at $r_E$. Note that the selected LTW at $r_F > r_A$ has a broad width $\approx 9.4$, while the LTW's selected below $r_A$ are narrow pulses like those at $\psi = -0.08$. Recent restricted test calculations for a $\Gamma = 40$ system, on the other hand, seem to indicate that there might be no stable LTW at $r_F$ in a system of length $\Gamma = 40$. There we performed two simulations: (i) While starting with a transient of width $20$, we got the LTW $C_2$ at $r_A$ after a *shrinking* process. But after increasing $r$ to $r_F$ having initially a length $\ell \approx 15$ this LTW transient *expanded* during $1200$ diffusion times to $\ell \approx 20$ before we stopped the run. (ii) Starting with the LTW $C_2$ at $r_A$, we increased the drive linearly to $r_F$ over $100$ diffusion times. During this time, $\ell$ increased from $9.6$ to about $10.5$. Then, we kept the drive at $r_F$, and the width still increased and grew to about $\ell \approx 16$ after $1000$ further diffusion times. The width $\Delta X$ increased sublinearly from $2.81$ to $4.27$ during this interval, which is well above the widths of the states $F_1, F_2$ and also $C_1, C_2$. We do not know whether there is a LTW attractor at $r_F$ in a $\Gamma = 40$ system or whether the expansion proceeds until there is an erratic interaction of the leading and trailing front as in the $\Gamma = 20$ system. Furthermore, data for $r$-values between $r_A$ and $r_F$ are not available to determine, e.g., the upper band limit of broad LTW states in a $\Gamma = 40$ system.

(f) *Multistability of different LTW's at $r_A$.* — Points (d) and (e) above already suggested that the vicinity of $r_A = 1.246$ is somewhat special, as we discuss now in more detail. Consider first the behavior in a $\Gamma = 20$ system. On the one hand, there is the narrow LTW $A_1$ that grew out of $D_1$ after increasing $r$ from $r_D$ to $r_A$. On the other hand, there are the broad states $B_1$ that evolved out of a somewhat broader LTW transient and the state $B_2$ — being practically identical to $B_1$ — which was generated with a completely different procedure: Starting from $B_1$, we reduced $r$ to $r_E$ but did not wait until state $E_1$ developed. Instead, we increased $r$ back to $r_A$ as soon as the transient LTW had reached an extension $\Delta X \approx 2.16$ between $A_1$ and $B_1$. Thereafter, the width grew to $B_2$. So with the two scenarios of a transient shrinking to $B_1$ and a transient growing to $B_2$, we have a selection of the state $B_1 = B_2$. In addition, there is the other stable state $A_1$ with a smaller width.

(g) *System length dependence.* — To test for length dependences, we compared runs for a $\Gamma = 20$ system with runs for a $\Gamma = 40$ system. Comparing the narrow monostable states $D_1$ in a $\Gamma = 20$ system and $D_2$ in a $\Gamma = 40$ system at $r_D$, below $r_A$, we did not find any difference: Not only the final-state properties but also the relaxation rates towards the final states were the same. On the other hand, at $r_F$, above $r_A$, our recent calculations discussed in (e) seem to argue against the existence of a stable LTW in a $\Gamma = 40$ system, while for $\Gamma = 20$ there is a monostable LTW state $F_1 = F_2$.

Now consider the situation at $r_A$. To test for length dependence of the two bistable states $A_1$ and $B_1 = B_2$ that we found at $r_A$ in a $\Gamma = 20$, system we did the following runs: First, we used $B_1$ as initial condition in a $\Gamma = 40$ system and observed an expansion (Fig. 7) to the slightly broader state $C_1$. So there is indeed a small $\Gamma$ dependence of the broad state at $r_A$ which is absent for the narrow states $D_1, D_2$ at $r_D$. In another run, we "filled" half of the $\Gamma = 40$ system with an extended TW and the other half with the conductive state. This initial condition shrank to a broad LTW transient that slowly relaxed with monotonically decreasing width (Fig. 7d) towards $C_2 = C_1$. These two scenarios show that we also have a state selection of $C_1 = C_2$ in $\Gamma = 40$ systems. Furthermore, the slow width-reducing relaxation from a transient that filled half of the $\Gamma = 40$ system towards $C_2$ suggests that $C_1 = C_2$ is the broadest stable LTW at $r_A$. But $C$ is not the only stable LTW in $\Gamma = 40$ systems. As for $\Gamma = 20$, there is multistability. This we showed with the generation of $A_2$ starting from $D_2$ at $r_D$ and increasing $r$ to $r_A$ (Fig. 7). $A_2$ is distinctly narrower (by 3.1) than $C$ and only marginally wider (by 0.2) than its $\Gamma = 20$ analog $A_1$.

While the final-state properties of our LTW's at $r_A$ like phase velocity, extension, group velocity, and field amplitudes do not differ much for the two different system lengths, there is a big difference in relaxation times: In our $\Gamma = 40$ system, the relaxation rate into one of the bistable states at $r_A$ was about 0.001, i.e., a factor of 15 smaller than the rate in the $\Gamma = 20$ system – not just a factor of about 4 as one might expect from diffusive processes. So we had to wait some thousand diffusion times. Below $r_A$ at $r_D$, on the other hand, there was no difference in the relaxation rate towards the narrow states $D_1 = D_2$ obtained for different $\Gamma$'s.

Since Kolodner [14,15] did not see bistability in his system of length $\Gamma = 82$, but rather states of several different widths for $1.335 \lesssim r \lesssim 1.338$ [14, Fig. 4], one wonders whether bistability might be a finite-size effect that vanishes in the limit of $\Gamma \to \infty$ or indeed already for large finite $\Gamma$. On the other hand, if by increasing the system size by another factor of two from our $\Gamma = 40$ to his $\Gamma = 82$, the relaxation time increases again by a factor of 15 to about 15000 one wonders whether such ultra slow dynamics could be changed by ultra small inhomogeneities.

(h) *Conclusion.* — So we found monostability or unique selection of a narrow (broad) LTW at $r_D, r_E$ ($r_F$) in a $\Gamma = 20$ system in the lower (upper) part of the $r$-band of LTW states. In a $\Gamma = 40$ system, we found unique selection of the same narrow LTW state as in the $\Gamma = 20$ system at $r_D$, i.e., in the lower part of the $r$-band. Above $r_A$, we have only test data available for a $\Gamma = 40$ system at $r_F$ which are not conclusive but hint against the existence of a broad LTW. In that case, the upper band limit of stable LTW's would be somewhere between $r_A$ and $r_F$ in the $\Gamma = 40$ system, i.e., below the band limit of the $\Gamma = 20$ system. Right at $r_A$, our results show, both in a $\Gamma = 20$ as well as in a $\Gamma = 40$ system, bistability of two different LTW attractors – a broad one and a narrow one. But the relaxation times towards them are significantly longer in the $\Gamma = 40$ system than in the $\Gamma = 20$ system. Thus, we speculate that (at least) the $\Gamma = 20$ data points shown by circles in Fig. 6b should be connected by a S-shaped curve which accomodates the monostable states on the upper and lower branches away from the turning-point region around $r_A$, as well as the pair of bistable states in between, as shown schematically in Fig. 6b. The $\Gamma$-dependence of $w_{max}, \ell, v_g$ (cf. Fig. 6), and of the relaxation time of the states seems to suggest that the $r$-subrange of bistability for the $\Gamma = 40$ system is somewhat narrower than for the $\Gamma = 20$ system. Lacking the resources to perform additional calculations, say, for $r$-values between $r_A$ and $r_F$, we cannot presently quantify the precise width of the bistability subrange in the two systems.

### B. Comparison with experiments

Very recently, Kolodner [14,15] has done experiments in an annular convection channel of periodicity length $\Gamma \approx 80$ with an ethanol-water mixture with parameters $L = 0.0079, \sigma = 8.93, \psi = -0.253$. A comparison with them shows some agreement, but also some differences with our calculations done with $\Gamma = 20$ and $\Gamma = 40$ for $\psi = -0.25$. Given that the experimental Prandtl number and in particular the Lewis number differs from our $\sigma = 10$ and $L = 0.01$, we will also include in our comparison his results for the next nearest separation ratio, $\psi = -0.210$ [15].

#### 1. Agreements

(i) The r-interval in which stable LTW's exist is very narrow – its width is of the order of $10^{-2}$.

(ii) LTW's drift with group velocities that are small compared to their phase velocity.

(iii) For small $r$-values, there is unique selection of a narrow pulse, the properties of which depend slightly on r. This can be seen in our calculations at $r_D$ and at $r_E$, in the experiment for $\psi = -0.253$ at $r = 1.334$ [14, Fig. 4], and in the experiment for $\psi = -0.210$ in the interval $1.327 \lesssim r \lesssim 1.335$ [15, Fig. 19].

(iv) For high $r$-values there is also a unique selection — however of a broad state, the properties of which depend on r. The selected width is significantly larger than the narrow width selected at lower $r$. The broader states are found at $r_F$ in our calculations with $\Gamma = 20$ and for $1.339 \lesssim r \lesssim 1.340$ in the experiment [14, Fig. 4].

(v) From the recent experiments [14,15], one can infer that the "arbitrary-width" LTW states that were observed earlier [4,5,9] in less uniform cells do not appear in more uniform systems. This agrees with our calculations.

(vi) Structural properties of the experimental LTW's [9,10,15], i.e., the topview shadowgraph intensity profiles and the wave number profiles [72], agree nicely with the numerical ones.

### 2. Differences

(i) The absolute r-values at which LTW's are found in the experiments at $\psi = -0.253$ and at $\psi = -0.210$ are larger than the calculated ones by about 8%. This, we think, is nothing to be worried about.

(ii) The smallest monostable narrow states that are reported for $\psi = -0.253$ have a width $\ell_{exp} \simeq 6.8$ versus $\ell_{num} \simeq 5.4$ for $\psi = -0.25$. Since we have not determined the width at the lower end of the r-band of stable LTW's, the smallest $\ell_{num}$ might even be slightly smaller, as can be inferred from the plot of $\ell(r)$ in Fig. 6b. Remember also the slightly different width of shadowgraph intensity and vertical velocity field, as discussed in Sec. IV. On the other hand, for $\psi = -0.210$ narrow pulses were observed [15, Fig. 19] in the r-interval $(1.327, 1.335)$ with widths $4.2 \lesssim \ell_{exp} \lesssim 5.9$ that are comparable to our $\ell_{num}$, but broad LTW's with widths up to 30 [15, Fig's. 15 and 18] were also seen in the r-range $(1.3355, 1.3385)$. Also, the broad monostable states for $\psi = -0.253$ in the r-interval $(1.339, 1.340)$ of [14, Fig. 4] have $\ell_{exp} \gtrsim 15$, while our broad, monostable states have $\ell_{num} \approx 10$. Furthermore, Kolodner observes stable LTW's of width $\ell_{exp} = 21.059$ and $\ell_{exp} = 32.72$ [15, Fig. 14 and 13]. Since the different definitions of LTW widths (cf., sec. IV.A for a discussion) via the profiles of the top-view shadowgraph intensity distribution and of the vertical flow intensity, respectively, differ here by at most $O(0.5)$, they cannot account for the width differences of *broad* experimental and numerical LTW's. However, we should mention that we cannot investigate very broad states in our $\Gamma = 20$ system and have not done it extensively in our $\Gamma = 40$ system. If very broad, stable LTW's, e.g., with length $15 \lesssim \ell \lesssim 30$ should exist in our $\Gamma = 40$ system, then they should exist between $r_A$ and $r_F$, and their r-range would have to be smaller than $\Delta r \approx 0.002$.

(iii) The experimental drift velocities for $\psi = -0.253$ are negative. They scatter around $-0.02$ [14, Fig. 2] but show no particular r-dependence. For the narrow experimental states of $\psi = -0.210$, they increase considerably from $-0.05$ to about 0 [15, Fig. 19b]. Our calculated group velocities of Fig. 6b, on the other hand, do not vary much. They increase only slightly with r, except for the bistable range, where the behaviour is somewhat more complicated. Furthermore, all our drift velocities are positive — around 0.05. This difference might have several causes: (1) The drift velocity with which leading and trailing fronts of the LTW move results from a complicated balance between stabilization and destabilization of conduction and convection in which concentration currents and buoyancy forces play an important role and which therefore is presumably quite sensitive to details like, e.g., the cell width. For example, too strong a current-induced stabilization of the conductive state ahead of the leading front (sec. III.B.4) should cause an invasion of the conductive state there into the convective region, i.e., a recession of the leading front. (2) The Lewis numbers in the experiment are smaller than in the simulation, and L is very important, at least for extended states — see also the discussion in sec. IV.A.1 on the effect of reducing L upon the drift velocity. (3) The calculations are 2d, while the experiments probe convection in an annular channel with a small channel width — see, for instance, the comparison of experimental and numerical LTW's by Surko et al. [9].

(iv) Kolodner [15] states, despite the scatter of the data [14, Fig. 4] for $1.335 \lesssim r \lesssim 1.338$, $\psi = -0.253$ that he always observes only one stable selected width. He also observes such monostability with less scatter for $\psi = -0.210$. This has to be contrasted with the bistability that we see in the narrow r-interval of Fig. 6a. Before concluding that there is indeed a discrepancy here, one should consider the scatter of the data, the smallness of the r region of bistability, and the question of system length dependence and long relaxation times mentioned above.

## VI. CONCLUSION

The main goal of this paper has been to provide a quantitative description of spatially-confined travelingwave convection in binary fluid mixtures and thereby to come to an understanding of these strongly nonlinear LTW states that compete with spatially extended convection for negative separation ratios in a narrow interval of Rayleigh numbers. Since the concentration field plays a decisive and very important role — after all, without it, i.e., in pure fluids there are neither TW's nor LTW's — the analysis of its influence on the balance of buoyancy forces that drive convection have turned out to be central to understanding the spatio-temporal behaviour of LTW's. Simply constructed, heuristic amplitudeequation models that do not incorporate the degrees of freedom of the concentration field appropriately should therefore not be expected to provide a proper description of LTW's.

We have numerically determined LTW solutions of the full, 2d field equations in laterally periodic systems of length $\Gamma = 20$ and $\Gamma = 40$ for ethanol-water mixtures ($L = 0.01, \sigma = 10$) with negative Soret coupling parameters $\psi = -0.08$ and $\psi = -0.25$, for which experiments

[2–15] have been performed.

*Structure.* — Our LTW's drift with a finite but small group velocity $v_g$ forwards, i.e., in the propagation direction of the phase of its TW constituents. In the frame comoving with velocity $v_g$, LTW states are time periodic. Remarkably enough, they are also symmetric under time translation by half an oscillation period combined with reflexion through the horizontal mid plane of the fluid layer. LTW's consist of three characteristic parts: a leading front (with respect to the phase propagation direction), a central part, and a trailing front. Roughly speaking, all our LTW states differ only in the width of the central part. Structural properties under the two fronts are universal but different for trailing and leading fronts. The amplitudes of velocity and convective temperature fields in the center of LTW's are slightly smaller than in the extended states at the same parameters. The LTW frequency, being about half the Hopf frequency at $r_{osc}$, is considerably bigger than that in the competing nonlinear extended state. Consequently, the LTW concentration contrast between adjacent rolls and the mean lateral concentration current is significantly larger than in the extended state.

The LTW profiles of the velocity, temperature, and concentration field have different shapes and widths and thus cannot be described by one common amplitude. The wavelength $\lambda$ of a particular roll pair, and with it its phase speed $v_p$, increases monotonically while it moves from its generation under the trailing front towards its death under the leading front. In the center, $\lambda$ is about 10% smaller than in the extended state. Otherwise, the fields in the center of LTW states have a similar structure as in extended TW states of similar frequency. This includes the mixing and boundary-layer behaviour of the concentration field with its characteristic plumes and trapezoidal lateral profiles.

*Buoyancy balance and concentration redistribution.* — Averaging over one oscillation period of the LTW, one finds that the convection-induced mean buoyancy force $\langle b \rangle = \langle c + \theta \rangle$ is dominated by the concentration contribution. The different widths of the pulses of $\langle c \rangle$ and $\langle \theta \rangle$ give rise to a relative enhancement of $\langle b \rangle$ right under both fronts that stabilizes the LTW against invasion of the conductive state. On the other hand, ahead of the leading front, a current-induced concentration redistribution or "barrier" is produced so as to weaken $\langle b \rangle$ there and thus to impede a rapid invasion of the quiescent conductive region by convection. The large-scale concentration redistribution is induced by a large-scale mean concentration circulation extending over the whole LTW. This mean concentration current is driven by a phase shift between the concentration wave and the velocity wave which occurs in the center part of the LTW, as in an extended TW. In the center part, the current is horizontal and of convective nature, and it flows in opposite directions in the upper and lower half of the layer. Under the fronts, the current loop is closed by vertical parts with gradient-related diffusive contributions.

With decreasing LTW frequency, i.e., with increasing $r$, the mean concentration current decreases and with it the concentration redistribution, the "barrier" ahead of the front, and the associated hindering of the forwards motion of the leading front. Thus, the forward drift velocity increases. Vice versa, for larger $\omega$, the current-induced, buoyancy-reducing vertical concentration distribution ahead of the leading front (the "barrier") becomes stronger and reduces more effectively the forwards drift velocity of the pulse into the conductive region. The experimentally observed reduction of $v_g$ below zero for LTW's with even larger $\omega$ also fits into this picture: the concentration distribution piled up by the fast waves — $\omega$ and with it the concentration current is large — at the leading edge is strong enough to stabilize conduction there and to push back the convective LTW region.

*Narrow pulses at small negative Soret coupling.* — For $\psi = -0.08$, we found LTW's of uniquely selected narrow width as in experiments [6,13,10,15]. The $r$-band in which these pulses are stable lies in the "convectively unstable" regime above the threshold $r_{osc}$ for growth of extended infinitesimal TW perturbations. So these states ultimately owe their stability to the finiteness of the system — in the "convectively unstable" regime wave packets of infinitesimal perturbations of the conductive state move with the fast critical group velocity and thus are absorbed in a finite system by the strong LTW state before they reach amplitude levels that can destroy the LTW pulse [21,22,17,10,12]. In an infinite system an infinite string of properly spaced pulses would be necessary. The stable LTW band also lies above the end point $r^*$ of the stable nonlinear extended TW solution branch, so that the LTW attractor coexists with the one of the extended SOC solution.

The $r$-variation of various properties of the pulses within the stable band is in good agreement with experiments by Kolodner [10,15] and Niemela et al. [6], if one ignores that the latter pulses [6] have zero drift velocity — presumably due [11,10] to some experimental pinning inhomogeneities. In particular, the LTW frequency decreases strongly with $r$, which has also been found for experimental pulses resting in a straight, rectangular convection channel [17] and for drifting pulses in an annular channel [15]. As an aside, we mention that the decay and temporal evolution of a pulse differs significantly after crossing the lower or upper boundary of the $r$-band for stable LTW's.

*LTW's at stronger negative Soret coupling.* — At $\psi = -0.25$, LTW's show richer and more complex bifurcation properties than at $\psi = -0.08$. Again, there is a finite, narrow band of $r$-values in which stable LTW's exist. Here, the band lies below $r_{osc}$, and the attractors of LTW, extended TW, and conductive state coexist therein with each other. Our results can be summarized and interpreted as follows: In the lower (upper) part of this band there is a uniquely selected narrow (broad) LTW state for each $r$, i.e., only one LTW attractor. But in between, there is bistability of two different LTW at-tractors with a narrow and a broad width, respectively. Our findings are compatible with an S-shaped curve $\ell(r)$ of LTW width $\ell$ versus $r$ that accomodates the monostable narrow and broad states on the lower and upper branch as well as the two bistable LTW attractors in the turning region of the curve $\ell(r)$. Our less extensive results for the system length $\Gamma = 40$ suggest that there (i) the upper band limit of stable LTW's lies below that of the $\Gamma = 20$ system and that (ii) the $r$-subrange of bistability is narrower than for $\Gamma = 20$. For a detailed quantitative comparison of our LTW solutions of the 2d hydrodynamic field equations with experimental properties of LTW convection in annular channels obtained recently [14,15], and a discussion of common and different properties, we refer to section V.B.

We finally mention that we found irregular behaviour when crossing the upper existence boundary of the LTW band: First the LTW pulse expanded into the surrounding conductive state — either at both fronts or with the leading front moving faster into the same direction as the trailing one. When, in our laterally periodic system, both fronts "touched" each other they interacted strongly and caused long time complicated irregular behaviour, e.g., with strong variations in the Nusselt number.

## ACKNOWLEDGMENTS

Discussions with G. Ahlers, K. D. Eaton, P. Kolodner, S. J. Linz, D. R. Ohlsen, H. Riecke, D. Roth, and C. M. Surko are gratefully acknowledged. We thank P. Kolodner for carefully reading the manuscript and for suggesting improvements. This work was supported by the Deutsche Forschungsgemeinschaft and the Stiftung Volkswagenwerk.

TABLE I. LTW states for $\psi = -0.08, L = 0.01, \sigma = 10$. The LTW state $X$ was already presented in [42,26].

| state | $X$ | $Y$ | $Z$ |
|---|---|---|---|
| $r$ | 1.104 | 1.106 | 1.109 |
| $\Gamma$ | 20 | 20 | 20 |
| $N-1$ | 0.0137 | 0.0155 | 0.0173 |
| $w_{\max}$ | 2.75 | 2.98 | 3.15 |
| $\breve{\omega}$ | 2.29 | 2.02 | 1.55 |
| $l$ | 4.8 | 4.9 | 5.0 |
| $\Delta X$ | 1.40 | 1.42 | 1.44 |
| $v_g$ | 0.052 | 0.066 | 0.083 |

TABLE II. LTW states for $\psi = -0.25, L = 0.01, \sigma = 10$. Due to improved evaluation techniques, e.g., for the frequency, some of the data of the final LTW states presented here differ slightly from our earlier estimates [42,26] for $A_1$, $B_1$, $C_1$.

| state | $D_1$ | $D_2$ | $E_1, E_2$ | $A_1$ | $A_2$ | $B_1, B_2$ | $C_1, C_2$ | $F_1, F_2$ |
|---|---|---|---|---|---|---|---|---|
| $r$ | 1.241 | 1.241 | 1.244 | 1.246 | 1.246 | 1.246 | 1.246 | 1.248 |
| $\Gamma$ | 20 | 40 | 20 | 20 | 40 | 20 | 40 | 20 |
| $N-1$ | 0.043 | 0.021 | 0.048 | 0.053 | 0.029 | 0.079 | 0.042 | 0.083 |
| $w_{\max}$ | 5.0 | 5.0 | 5.06 | 5.17 | 5.08 | 5.25 | 5.18 | 5.28 |
| $\breve{\omega}$ | 4.38 | 4.37 | 4.31 | 4.33 | 4.38 | 4.35 | 4.46 | 4.34 |
| $l$ | 5.4 | 5.4 | 5.9 | 6.3 | 6.5 | 9.0 | 9.6 | 9.4 |
| $\Delta X$ | 1.50 | 1.49 | 1.61 | 1.74 | 1.84 | 2.43 | 2.59 | 2.54 |
| $v_g$ | 0.042 | 0.042 | 0.050 | 0.054 | 0.053 | 0.052 | 0.048 | 0.056 |

FIG. 1. Localized and extended convective states in ethanol-water mixtures ($L = 0.01, \sigma = 10$) in the bifurcation diagrams of frequency $\omega$ and of maximal vertical flow velocity $w_{max}$ vs Rayleigh number for two different separation ratios $\psi$. Symbols represent numerically obtained stable states. Curves are fitted splines to guide the eye. Schematic dashed curves for unstable lower branches show the bifurcation topology of TW's [79] and of SOC's [80]. For $\psi = -0.08, -0.25$, the oscillatory bifurcation threshold is at $r_{osc} = 1.0965, 1.3347$; Hopf frequency $\omega_H = 5.753, 11.235$; TW saddle $r^s_{TW} \simeq 1.06, 1.215$; transition $TW \longleftarrow SOC$ $r^* \simeq 1.09, 1.65$. Properties of the LTW states $X, Y, Z$ and $D_1, E, A_1, B, F$ shown here are listed in table I and II and are discussed in sec. IV and V, respectively.

FIG. 2. Comparison of the narrow LTW $A_1$, for $\psi = -0.25$ (left column), and the LTW $X$, for $\psi = -0.08$ (right column). Both are presented as traveling to the left. (a) Wavelength $\lambda(x)$ in the center part of the LTW with sufficient convective intensity. $\lambda(x) = \breve{v}_p(x) 2\pi/\breve{\omega}$ was determined from the frequency $\breve{\omega}$ and the phase velocity $\breve{v}_p(x)$ of the nodes of the vertical velocity field at $z = 0.5$ in the frame comoving with the drift velocity of the LTW and was checked against the node distances. (b) Snapshot (thin line) and envelope (thick line) of vertical velocity field $w(x, z = 0.5)$. (c) Streamlines of the time-averaged concentration current $\langle \mathbf{J} \rangle$. (d) Lateral profile of the time-averaged temperature field $\langle \theta \rangle$, concentration field $\langle c \rangle$, and the buoyancy force $\langle b \rangle = \langle c + \theta \rangle$ at $z = 0.25$. For better resolution, the right column for $\psi = -0.08$ shows $2w$ in (b) and $3\langle \theta \rangle$, $3\langle c \rangle$, and $3\langle c + \theta \rangle$ in (d). Profiles at $z = 0.75$ are related by the symmetry (3.6).

FIG. 3. Comparison of LTW and TW for common parameters. Both are presented as traveling to the left. The LTW ($A_1$) is the one of Fig. 2 (left column), here, however, presented at a different time, so that, e.g., the node positions of $w$ are different. The TW is the state of [46, Fig. 3b], but here traveling to the left. The first row shows the colour-coded temperature field $\delta T$ with arrows denoting the velocity field $\mathbf{u}$ of the LTW. The second and last row show the sideview shadowgraph intensity $I(x, z)$ of both states. The concentration field of both states is presented in the fourth and seventh row. The LTW's time-averaged concentration field (colour), together with streamlines (broken curves) of the mean concentration current, is presented in the third row. All concentration pictures use the same colour code. In the fifth and sixth row, lateral profiles of $w$ (thin line), $30\,\delta T$ (triangles), and $150\,\delta C$ (squares) are given at midheight, $z = 0.5$.

FIG. 4. $r$-dependence of narrow LTW states at $\psi = -0.08$. (a) Frequency $\breve{\omega}$, (b) maximum $w_{max}$ of the vertical velocity field, (c) width $\ell$ defined by FWHM of intensity envelope of vertical velocity field $w$ at midheight of the layer. (d) group velocity $v_g$. See also table I.

FIG. 5. Time evolution into the LTW $X$ at $r = 1.104$, $\psi = -0.08$. (a) frequency $\breve{\omega}$, (b) Nusselt number $N - 1$, (c) group velocity $v_g$, (d) standard deviation $\Delta X$ (4.2). Since frequency data were not recorded in the time interval between 200 and 220 we show in (a) a linear interpolation for this interval.

FIG. 6. $r$-dependence of narrow (open symbols) and wide (filled symbols) LTW states at $\psi = -0.25$. (a) Maximum $w_{max}$ of the vertical velocity field, (b) FWHM of vertical flow intensity $l$, (d) group velocity $v_g$. States in a system of periodicity length $\Gamma = 20$ are given as circles and states in a system of length $\Gamma = 40$ as triangles. Line through the $\Gamma = 20$ states is a guide to the eye. The thin S-shaped part is schematic to indicate the bifurcation topology discussed in the text. See also Tab. II for an identification of the different states.

FIG. 7. Time evolution into LTW's at $r = r_A = 1.246, \psi = -0.25$ in a system of periodicity length $\Gamma = 40$. (a) frequency $\breve{\omega}$, (b) Nusselt number $N - 1$, (c) group velocity $v_g$, (d) standard deviation $\Delta X$.

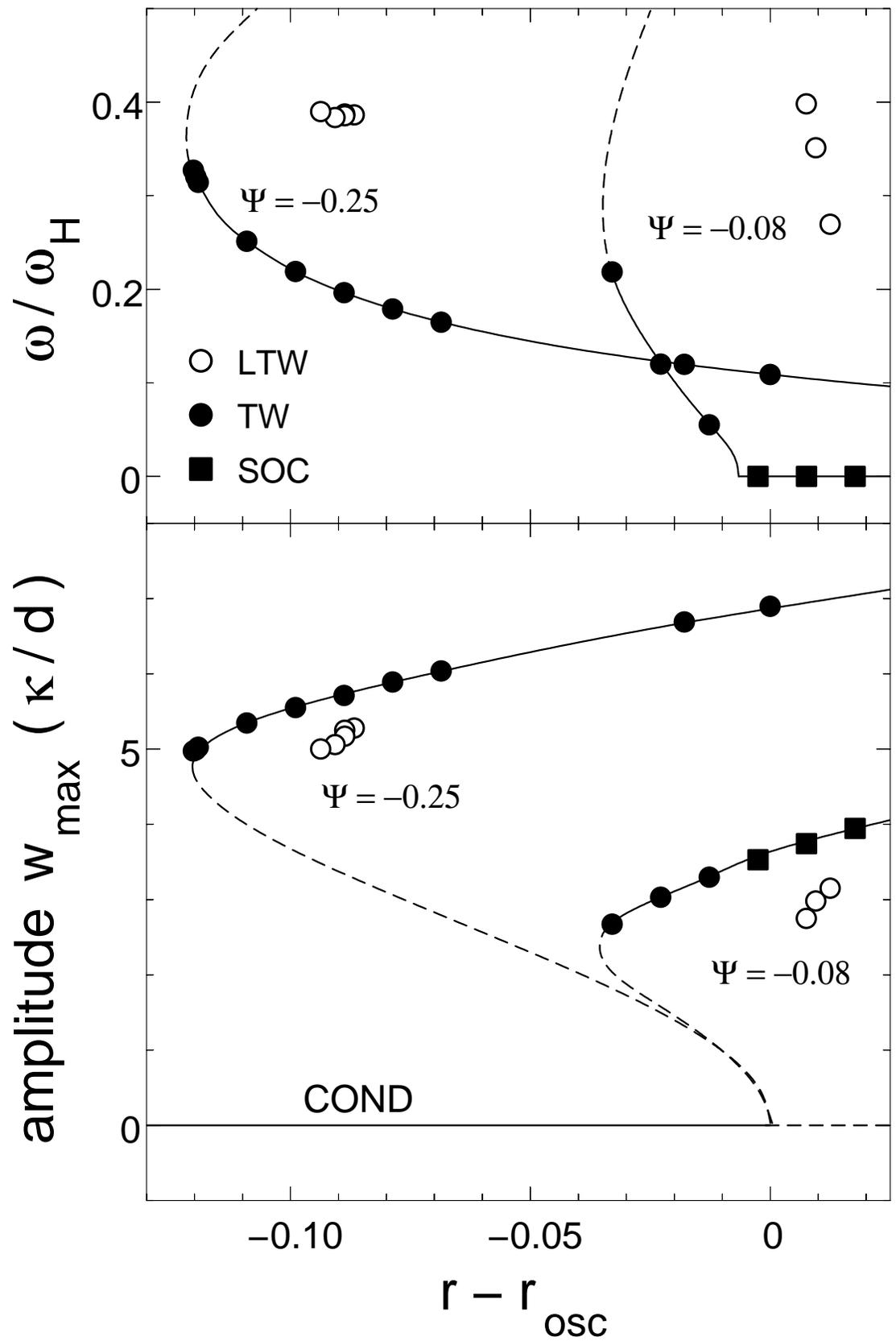

Fig. 1
Barten et al., Convection in Binary Fluid Mixtures II, Phys. Rev. E

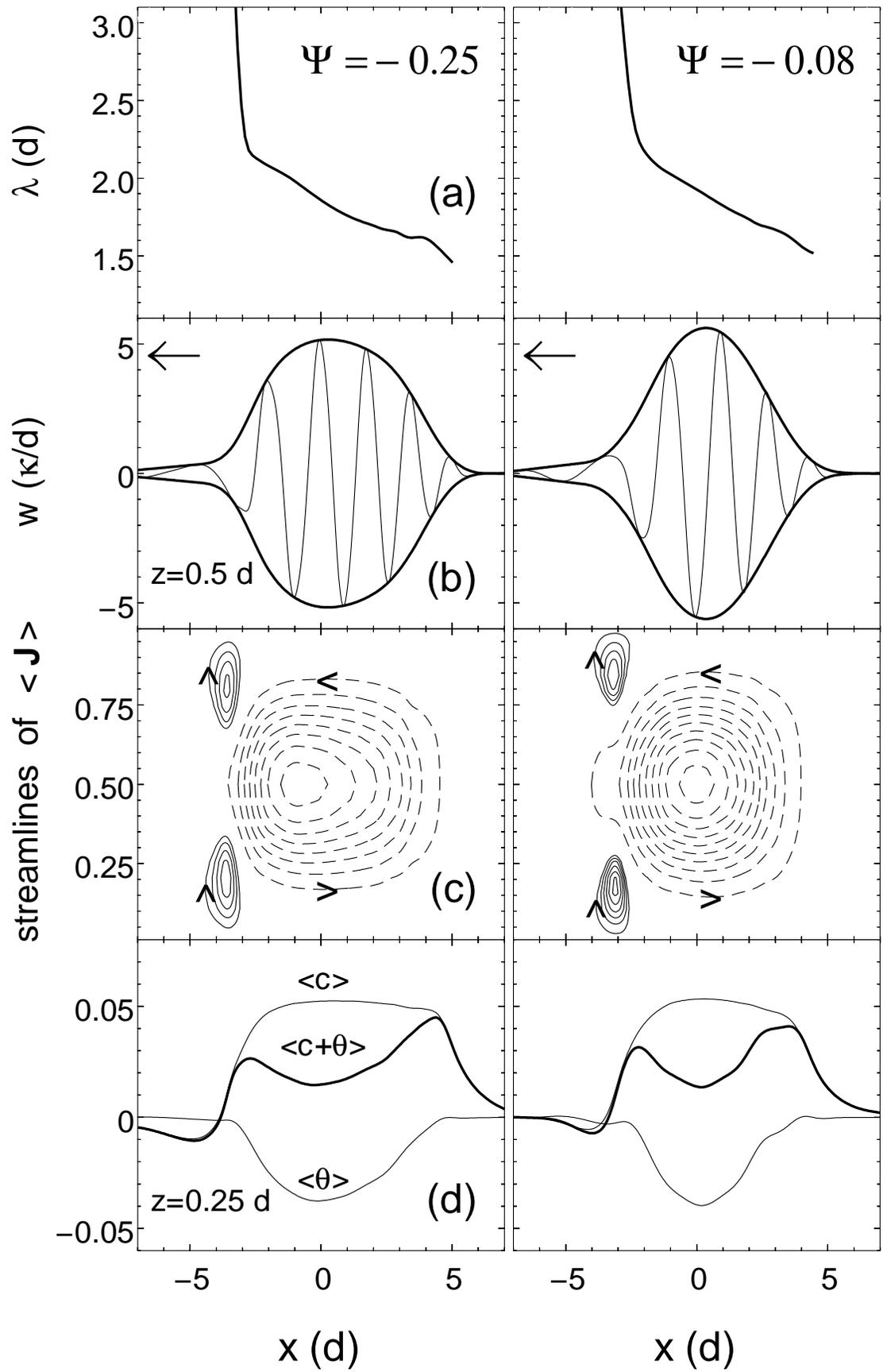

Fig. 2
Barten et al., Convection in Binary Fluid Mixtures II, Phys. Rev. E

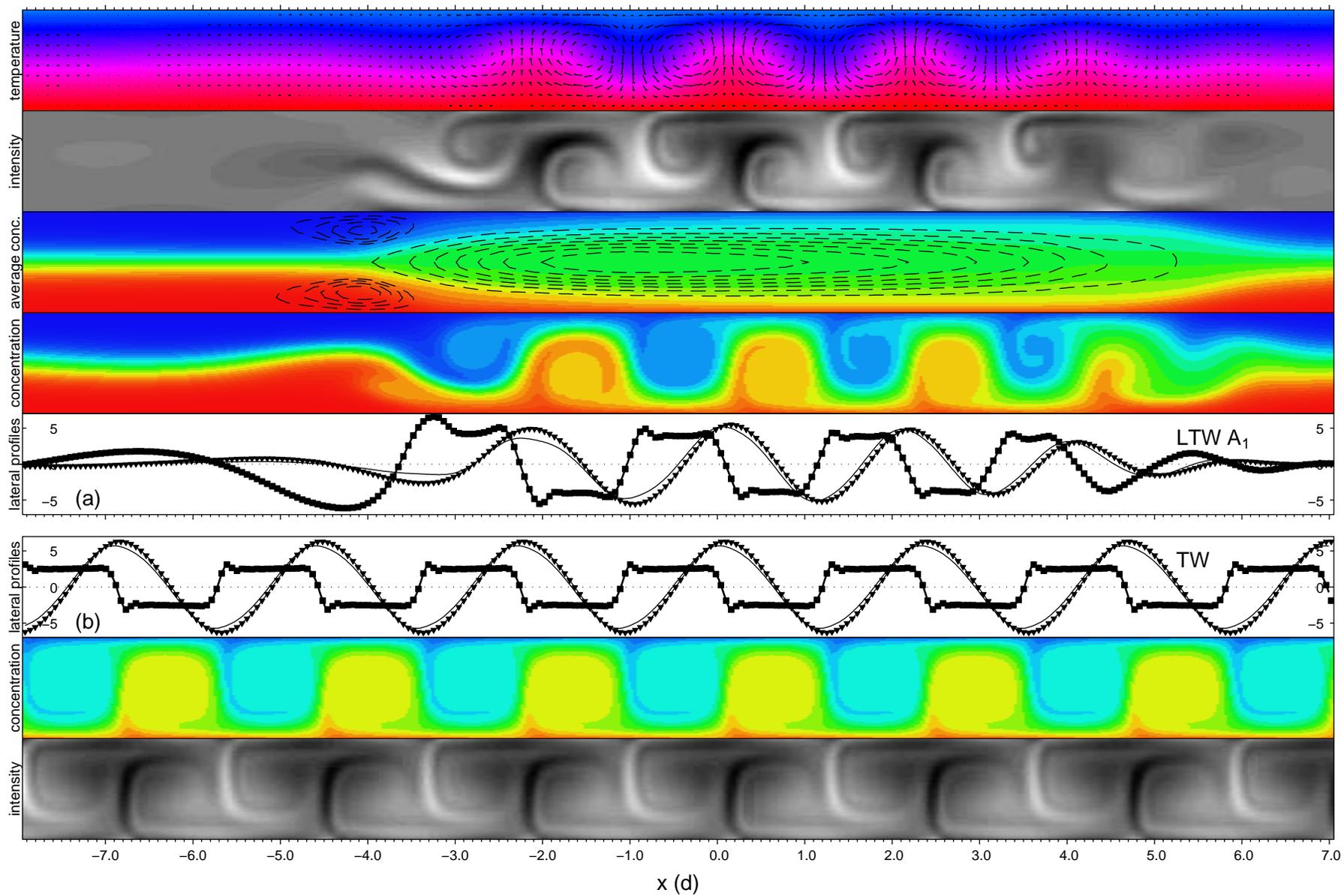

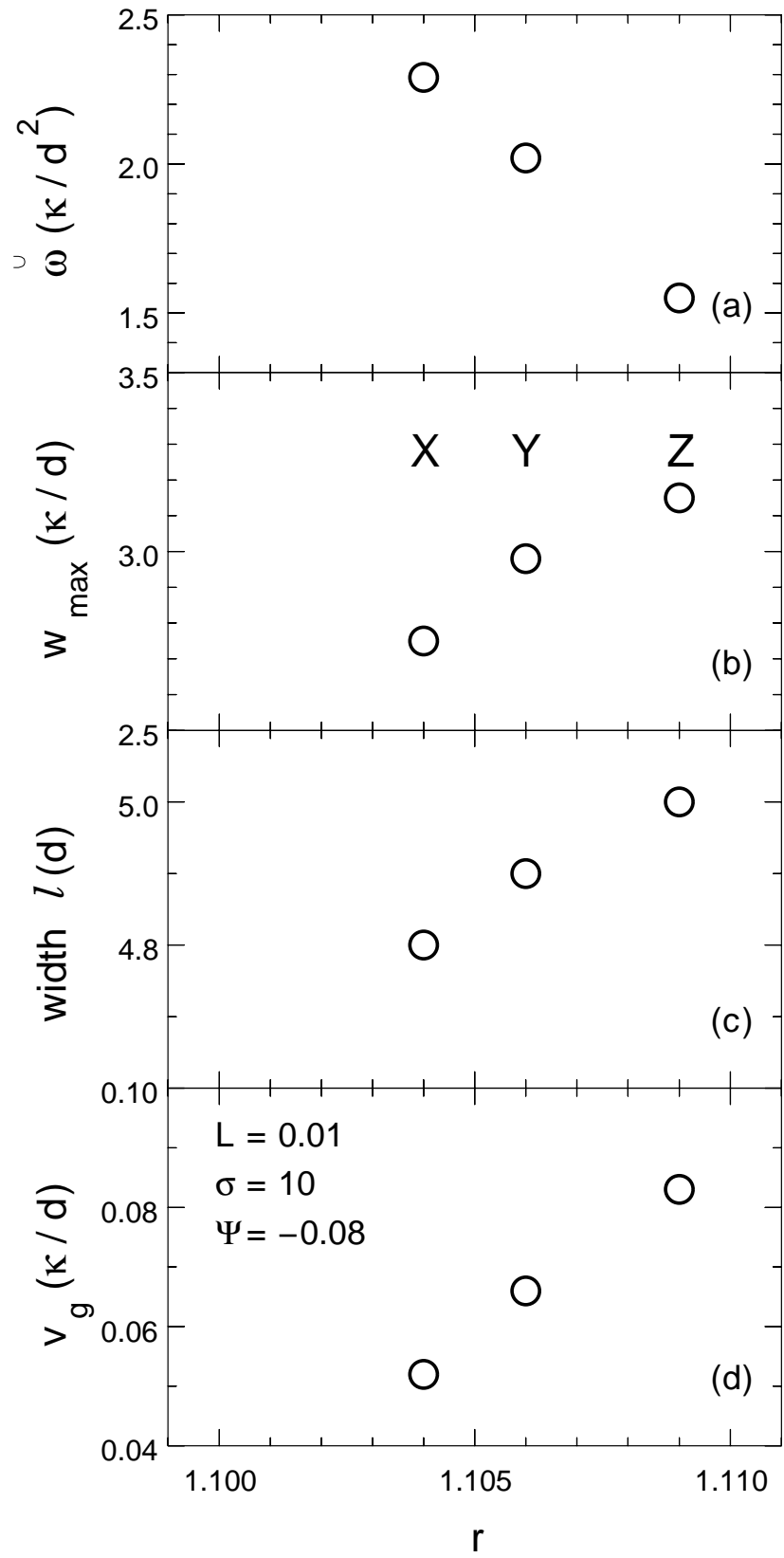

Fig. 4
Barten et al., Convection in Binary Fluid Mixtures II, Phys. Rev. E

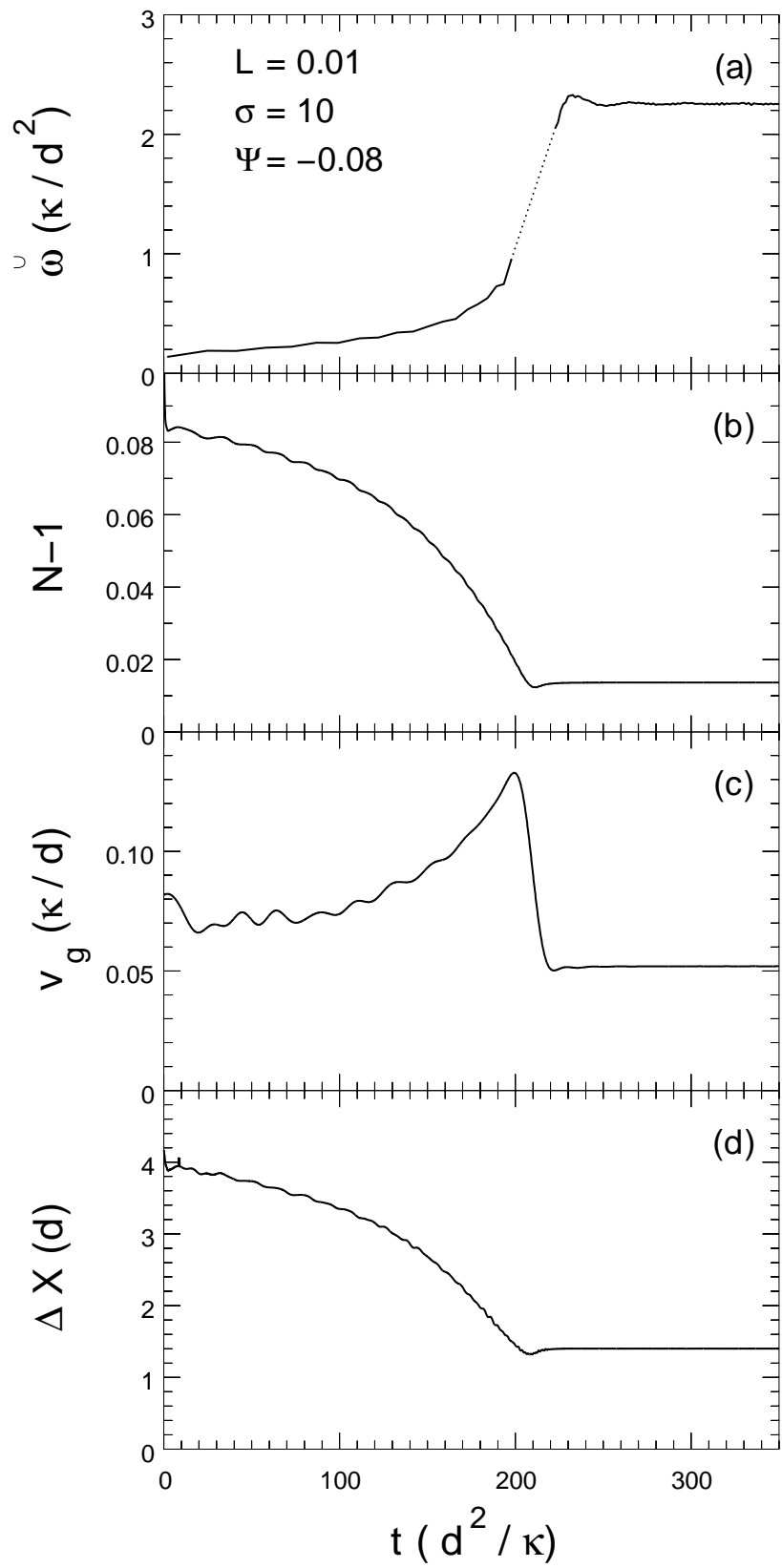

Fig. 5
Barten et al., Convection in Binary Fluid Mixtures II, Phys. Rev. E

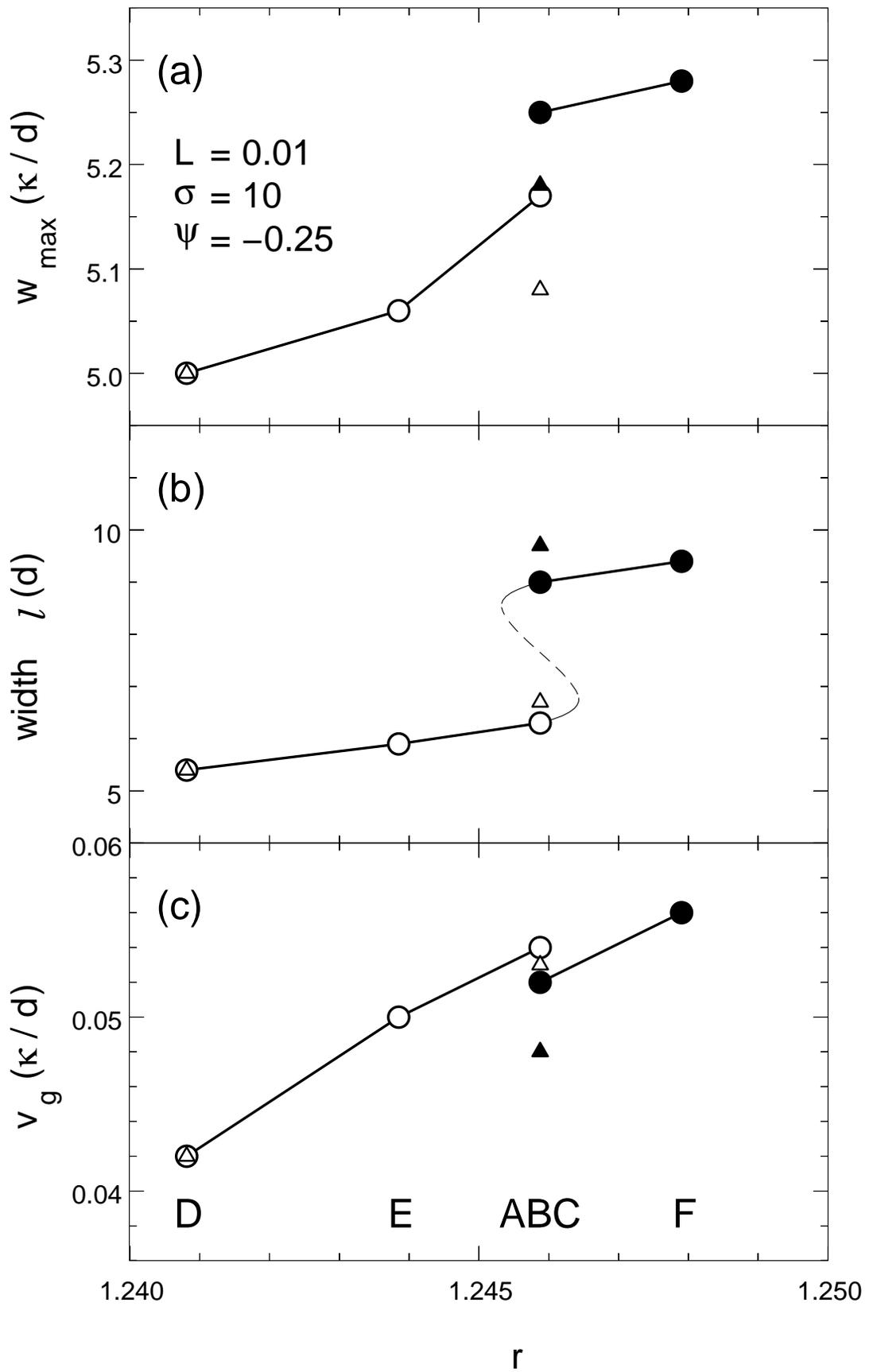

Fig. 6
Barten et al., Convection in Binary Fluid Mixtures II, Phys. Rev. E

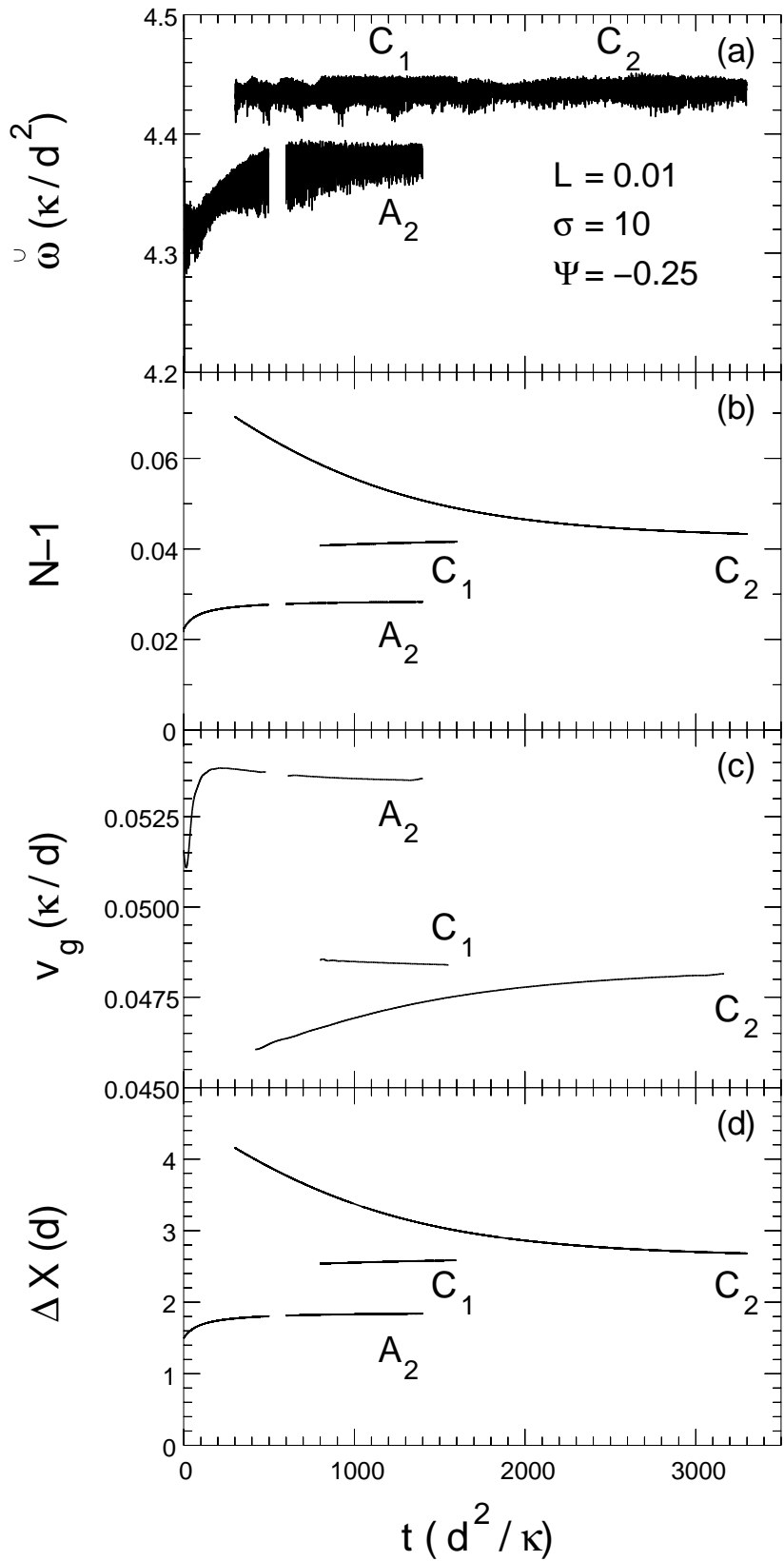

**Fig. 7**
Barten et al., Convection in Binary Fluid Mixtures II, Phys. Rev. E